\UseRawInputEncoding
\documentclass[%reprint,
superscriptaddress,
bibnotes,
 amsmath,amssymb,
 aps,
pra,
twocolumn,
longbibliography
]{revtex4-1}
\usepackage{graphicx}% Include figure files
\usepackage{dcolumn}% Align table columns on decimal point
\usepackage{xr}
\usepackage{bm} %bold math
\usepackage{float}
\usepackage{hyperref} %add hyperref capabilities
\hypersetup{
	colorlinks=true,
	linkcolor=blue,
	citecolor=blue,
	filecolor=magenta,
	urlcolor=blue         
}
\usepackage{color,soul}
\usepackage{lineno} %linenumbers

\renewcommand{\selectlanguage}[1]{}

\begin{document}

\title{Noise dynamics in large mode volume Brillouin lasers}

\author{Andrew J. Shepherd}
\email[]{as5262@nau.edu}
\affiliation{Department of Physics, Northern Arizona University, Flagstaff, Arizona 86011, USA}
\affiliation{Center for Materials Interfaces in Research and Applications, Northern Arizona University, Flagstaff, Arizona 86011, USA}

\author{Daniel J. Blumenthal}
\affiliation{Department of Electrical and Computer Engineering, University of California Santa Barbara, Santa Barbara, CA, USA}

\author{Ryan O. Behunin}
\email[]{ryan.behunin@nau.edu}
\affiliation{Department of Physics, Northern Arizona University, Flagstaff, Arizona 86011, USA}
\affiliation{Center for Materials Interfaces in Research and Applications, Northern Arizona University, Flagstaff, Arizona 86011, USA}

\date{\today}

\begin{abstract}
Photonic integrated Brillouin lasers have emerged as an important tool to realize a wide range of precision applications, including atomic time-keeping, low-noise microwave signal generation, fiber and quantum sensing, and ultra-high capacity coherent communications. While Brillouin lasers routinely achieve sub-Hz instantaneous linewidths, many of these applications also require exceptional frequency stability and high-power single-mode emission. A recent demonstration showed that extending the resonator length increases the laser power while simultaneously improving the frequency stability through suppression of low-frequency thermorefractive noise. However, as the resonator length scales to larger size, multiple optical resonances can be found within the Brillouin gain bandwidth, greatly complicating the laser dynamics and extending beyond the validity of existing coupled-mode Brillouin laser models. Given the potential to scale lasers of this type to watt-level output powers at sub-mHz linewidths, a theoretical model describing this physics is needed to provide key insights into their performance. Here, we develop a coupled-mode theory of integrated large mode volume Brillouin lasers, accounting for multiple cavity modes with potential to lase within the gain bandwidth. We obtain expressions for the steady-state dynamics, spontaneous spectrum, relative intensity noise, and frequency noise. Our analysis reveals that the broad gain bandwidth results in atypical Brillouin dynamics, giving rise to distinct features in the noise spectra, and consequently modifications of the standard, single-mode fundamental linewidth of Brillouin lasers. Additionally, these features may be used for a variety of tangential applications, such as phonon spectroscopy or quality factor enhancement. Furthermore, we find that the linewidth can be significantly impacted by transferred RIN from the external pump in Brillouin lasers that lack ideal phase matching.

\end{abstract}

\maketitle

\section{Introduction}
Integrated photonic lasers with low linewidth and high output power play a significant role in a wide range of precision applications, such as coherent optical communications \cite{kikuchi_fundamentals_2016}, atomic clocks \cite{ludlow_optical_2015}, environmental and quantum sensing \cite{mecozzi_use_2023, li_microresonator_2017, xu_sensing_2019}, coherent ranging (LiDAR) \cite{lihachev_low-noise_2022}, and coherent microwave and mmWave generation \cite{sun_integrated_2024, kudelin_photonic_2024}. Further reduction of linewidth and higher output power is critical to these applications, and also may unlock new opportunities where experimental reliability and portability is needed. Demonstrating narrow linewidth and high output power simultaneously has proved challenging; however, recent work has merged the effects of nonlinear feedback via stimulated Brillouin scattering with the noise-suppressing benefits of large mode volume optical resonators, achieving a fundamental linewidth of 31 mHz with 41 mW output power \cite{liu_large_2025}. This work suggests the potential for linewidth reduction to the sub-mHz level while maintaining watt-level output powers by increasing the resonator length; however, a coupled-mode theoretical framework of this system has not yet been described. Analyzing the unique underlying physics of this system may enable performance optimization, reveal fundamental limitations, and uncover potential new applications for this class of high performance lasers.

External cavity lasers (ECLs), employing semiconductor emission, leverage a large total intra-cavity photon number and long photon lifetime for narrow linewidths and high output power \cite{bai_comprehensive_2022}. State-of-the-art integrated ECLs can realize fundamental linewidths of $\sim10$ Hz and mW to 10s mW class output powers, though further improvements to either parameter involves a trade-off with the other  \cite{fan_hybrid_2020, heim_hybrid_2025, wu_hybrid_2024, morton_high-power_2018}. Given these limitations, a key motivation for advancing this class of lasers is to improve optical pump sources for stimulated Raman and stimulated Brillouin lasers \cite{fan_hybrid_2020, pahlavani_linewidth_2025, gundavarapu_sub-hertz_2019}, which offer access to even lower linewidths. Further reduction to integral linewidths in ECLs--both in lab-scale and integrated systems--can be realized by locking to a temperature-stabilized reference cavity \cite{matei_15_2017, liu_36_2022}, where the thermorefractive noise (TRN) floor is lowered due to the large reference cavity mode volume; however, a trade-off occurs between linewidth and high-frequency noise, as the reference cavity acts as a low-pass filter for phase noise beyond its linewidth, e.g., 100 kHz \cite{krinner_low_2024} and the feedback control actually amplifies higher frequencies \cite{liu_large_2025}.

High-performance lasing can be achieved with stimulated Brillouin scattering (SBS) lasers and self-injection locking (SIL) lasers, where nonlinear feedback is leveraged for noise suppression. The SIL laser suppresses noise via optical feedback, facilitated by a wavelength-selective element that filters light back into an already lasing semiconducting resonator \cite{lu_emerging_2024}. This technology has realized sub Hz linewidths in state-of-the-art systems \cite{li_reaching_2021, isichenko_sub-hz_2024}; however, attaining high output power without compromising linewidth remains a nontrivial task. SBS lasers use an entirely different form of nonlinear feedback, where laser-generated phonons suppress noise while mediating light scattering from an input mode (pump) to an output mode of slightly lower frequency (Stokes). This mechanism has been shown to inhibit pump noise transfer \cite{debut_experimental_2001, smith_narrow-linewidth_1991} and limit relative intensity noise (RIN) \cite{stepien_intensity_2002,molin_experimental_2008, geng_pump--stokes_2007}, where the phonon noise functions analogously to the spontaneous emission in conventional lasers resulting in analogies to the linewidth enhancement factor \cite{li_characterization_2012}. Along with the aforementioned laser designs, SBS laser development has been challenged with simultaneous narrow linewidth and high output power. Both linewidth and output power are generally hindered by cascaded emission \cite{behunin_fundamental_2018}, and although this effect can be canceled with specialized device design, output power grows with the square root of the injected pump \cite{puckett_higher_2019, liu_integrated_2024}. Additionally, the requirement for single mode lasing has driven designs in which the cavity length gives a free spectral range such that overlap occurs between only a single cavity mode and the Brillouin gain \cite{lee_chemically_2012, li_characterization_2012, gundavarapu_sub-hertz_2019}, ultimately limiting the optical mode volume and thus the TRN floor \cite{gorodetsky_fundamental_2004, huang_thermorefractive_2019}. Leveraging state-of-the-art fabrication techniques, an SBS laser designed in this manner realized a 245 mHz fundamental linewidth with 126 mW of output power \cite{qin_high-power_2022}.

Recent work, utilizing a 4 meter coil waveguide resonator, reaches a fundamental linewidth of 31 mHz and output power 41 mW \cite{liu_large_2025}, marking a new state-of-the-art design for simultaneously achieving narrow linewidth and high output power. The extended resonator length increases the intra-cavity photon number, reduces the TRN floor, and enables higher output power. Additionally, this system departs from previous approaches by enabling overlap between multiple cavity modes and a broad bulk-like Brillouin gain spectrum. Despite spontaneous scattering between multiple optical modes, this large mode volume system results in a single mode laser, where the cavity mode supporting the greatest Brillouin gain will lase. Furthermore, this work indicates that increasing resonator length may allow sub mHz fundamental linewidths while eclipsing 1 W of output power. 

Here, we develop a theoretical coupled-mode model to analyze the steady-state, spontaneous, and lasing dynamics of large mode volume Brillouin lasers. We model the broad Brillouin gain bandwidth with multiple-oscillators, in which each oscillator couples the pump to each cavity mode within the gain bandwidth. By assuming the acoustic fields decay rapidly in comparison to the optical fields, we find a set of nonlinear equations that describe the coupled dynamics of the pump and a collection of cavity modes. Solving the coupled steady state equations for the non-fluctuating amplitudes of each field, we find that the cavity mode with the highest Brillouin gain out-competes the other cavity modes within the Brillouin gain bandwidth, and begins to lase when its gain equals the loss. At this point, the pump amplitude along with all cavity modes of lesser Brillouin gain, clamp at a fixed steady state amplitude. We then linearize these equations around the steady-state for small amplitude perturbations, giving a compact set of equations describing the dynamics of the amplitude and phase fluctuations that lead to the RIN and frequency noise spectra. 

The broad Brillouin gain bandwidth gives rise to an inherent phase-mismatch, resulting in frequency pulling and linewidth narrowing of the cavity modes below threshold, pulling of the pump frequency above threshold, and unique features in the noise spectra at high offset-frequencies which results in a modification to the well-known SBS Schawlow-Townes-like linewidth \cite{li_characterization_2012, loh_noise_2015, debut_linewidth_2000}. Using these unique features, we find that this system may be used for non-lasing applications as well. The magnitude of the frequency pulling and narrowing of linewidth in each mode is significant enough to be detectable, and their measurement can lead to construction of the entire phonon gain spectrum. Additionally, the narrowed linewidth of the non-lasing cavity modes may permit ultra-high effective quality factors, increasing with resonator length as cavity modes become more closely spaced and their respective gains approach that of the lasing mode.

\section{Theory}
Conventional stimulated Brillouin lasers leverage photon-phonon coupling, where incident light scatters from an acoustic wave to a lower, backward propagating optical field. This interaction allows a high-frequency `pump' photon, with frequency $\omega_p$ and wavevector ${\bf k}_p$, to decay into a `Stokes' photon of lower frequency and a phonon with frequencies $\omega_s$, $\Omega$ and wavevectors ${\bf k}_s$ and ${\bf q}$ respectively. If phase matching is satisfied, the frequency relation $\omega_p=\omega_s+\Omega$ holds (conservation of energy) and similarly for the wave vectors, ${\bf k}_p={\bf k}_s+{\bf q}$ (conservation of `momentum'). Drawing analogies to traditional laser systems, the phonon's fluctuations facilitate spontaneous emission from pump-to-Stokes, and Stokes photons in turn stimulate the scattering process, ensuring optical amplification. To satisfy phase matching relations, resonators are normally designed to support a single cavity mode within the phonon gain bandwidth, and the cavity free spectral range (FSR) is fabricated to match the Brillouin shift. The amplification then occurs in a narrow gain bandwidth that depends on the decay rate $\Gamma$ of the acoustic field.
\begin{figure}
    \centering
    \includegraphics[width=8.6cm]{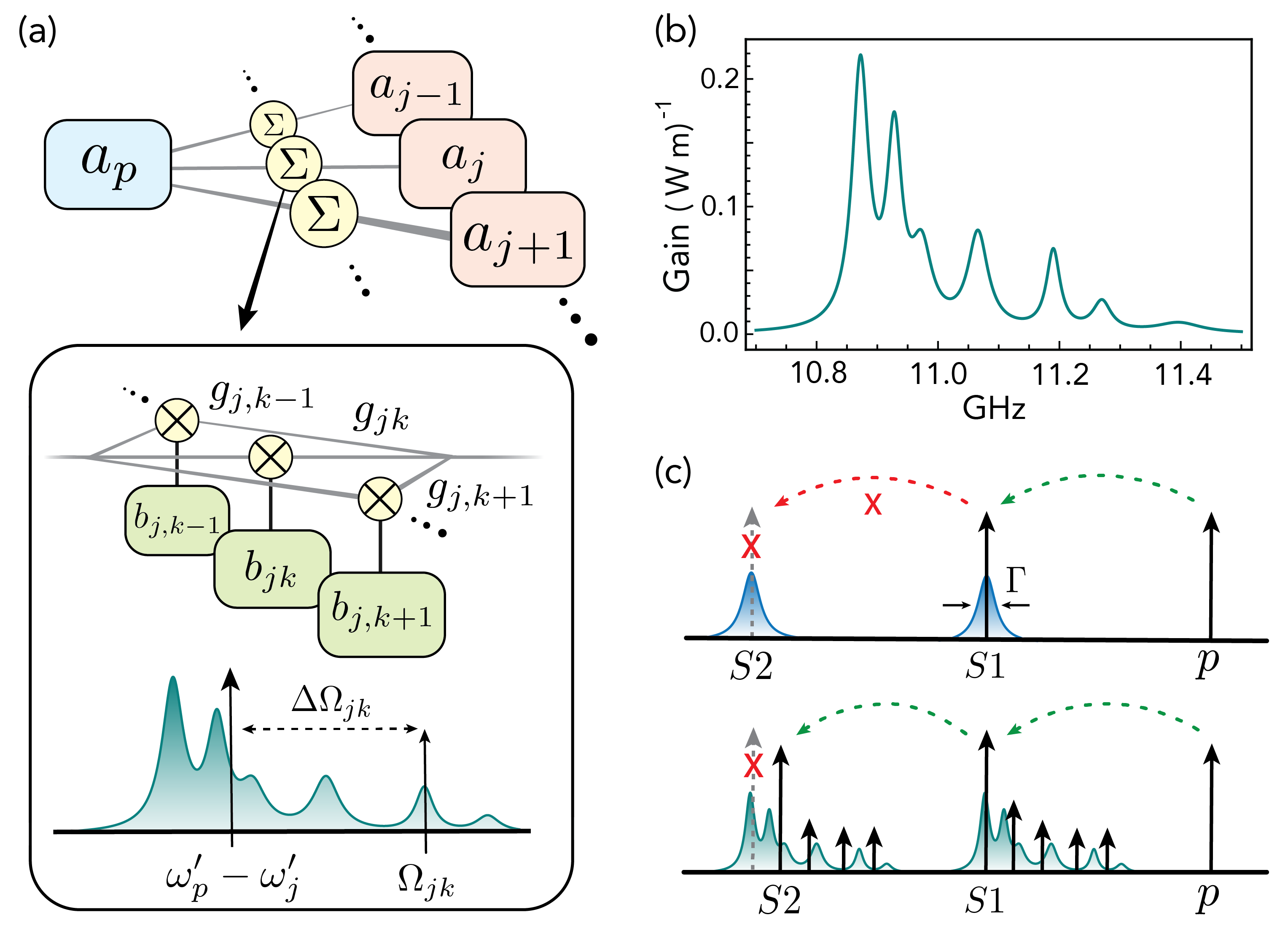}
    \caption{(a) Illustration of large mode volume Brillouin laser dynamics, where the pump mode $a_p$, is coupled (denoted by the sum and mixer symbol) to multiple cavity modes $a_j$, with each $a_p\rightarrow a_j$ coupling mediated by a distinct set of phonon modes $b_{jk}$. (b) Brillouin gain spectrum for a given cavity mode $j$. (c) Contrasting cascaded control in standard Brillouin lasers (top) vs. a large mode volume Brillouin laser (bottom), where removal of the lasing mode from the supported cavity modes will not cancel cascading.} 
    \label{fig:diagram}   
\end{figure}

In large mode volume Brillouin lasers, a greater resonator length supports a relatively small FSR, leading to multiple closely spaced cavity modes falling under a broad gain bandwidth. A pump laser will spontaneously scatter to each of these modes, mediated by a set of phonon modes in which each wavevector is distinct with respect to each cavity resonance. We model this interaction with the following Hamiltonian,
\begin{align}
    H=\hbar\omega_p a_p^\dag a_p+\sum_j\Bigl(\hbar\omega_j&a_j^\dag a_j+\sum_{k}(\hbar\Omega_{jk}b_{jk}^\dag b_{jk}\label{eq:H}\\&+g_{jk}a_pa^\dag_j b_{jk}^\dag + \rm{H.c})\Bigr),\nonumber
\end{align}
where $j$ indexes the cavity modes within the gain bandwidth, with annihilation operator $a_j$ and frequency $\omega_j$. The collection of phonon modes coupled to cavity mode $j$ are indexed by $jk$, with annihilation operator $b_{jk}$ and frequency $\Omega_{jk}$. The pump mode is represented with annihilation operator $a_p$ and frequency $\omega_p$, and H.c stands for the Hermitian conjugate. The interactions described in this Hamiltonian are illustrated in Fig. \ref{fig:diagram}(a), where the $\Sigma$ symbol denotes coupling between the pump and cavity modes. This symbol is further described by the lower box, showing a collection of mixer symbols which depicts how each $a_p\rightarrow a_j$ coupling is mediated by a collection of phonon modes. By modeling the system with Eq. \eqref{eq:H}, we are assuming the rotating wave approximation is valid, amplitudes are independent of space \cite{dallyn_thermal_2022, behunin_fundamental_2018}, and processes produced by Kerr nonlinearities, i.e., self- and cross-phase modulation, are negligible due to the relatively strong electrostriction effect in participating materials \cite{dallyn_thermal_2022}. The multi-phonon model can be described with the gain spectrum \cite{kharel_noise_2016},
\begin{align}
    G_{B,j}(\Omega)=\sum_{k}\frac{4|g_{jk}|^2L}{v_p v_j\hbar \omega_p \Gamma_{jk}}\frac{(\Gamma_{jk}/2)^2}{(\Gamma_{jk}/2)^2+(\Omega-\Omega_{jk})^2},
    \label{eq:gain}
\end{align}
shown in Fig \ref{fig:diagram} (b), where $\Gamma_{jk}$ is the decay rate for each phonon mode, $v_{p/j}$ is group velocity for each mode, and $L$ is the resonator length. All parameters can be found in Tables \ref{table1} and \ref{table2}, taken from either listed values in a physically realized system \cite{liu_large_2025}, or determined by finite element simulations using those values. The parameters comprising the phonon gain spectrum, in Table \ref{table2}, are found to be insensitive to cavity mode index $j$, and are therefore assumed constant with respect to $j$.

We derive equations of motion for the large mode volume laser using Eq. \eqref{eq:H} and adding in the effects of optical and acoustic damping, quantum and thermal fluctuations, and the external pump laser. We move to a rotating reference frame where $b_{jk}\rightarrow b_{jk}\exp\{-i\Omega_{jk}'t\}$, with $\Omega_{jk}'=\Omega_{jk}-\Delta\Omega_{jk}$ and $\Delta\Omega_{jk}$ is the difference between the driving beat note and the resonant phonon frequency, i.e., $\Delta\Omega_{jk}=\Omega_{jk}-(\omega'_p-\omega'_j)$ (visually shown in  Fig. \ref{fig:diagram}(a)). Additionally, we assume the pump mode and external laser are locked, oscillating at $a_p\rightarrow a_p\exp\{-i\omega_p't\}$, $F\rightarrow F\exp\{-i\omega'_pt\}$, and the cavity modes oscillate at $a_j\rightarrow a_p\exp\{-i\omega_j' t\}$. The associated frequencies are taken as $\omega_p'=\omega_p-\Delta\omega_p$ and $\omega_j'=\omega_j-\Delta\omega_j$, which accounts for the possibility of oscillation away from resonance. In doing this, the set of coupled Heisenberg-Langevin Eqs are
\begin{align}
    \dot{a}_p=-(i\Delta\omega_p&+\gamma/2)a_p-i\sum_{jk}g_{jk}^*a_jb_{jk}\label{eq:p_HL}\\&+\sqrt{\gamma_{ext}}F+\eta_p,\nonumber
\end{align}
\begin{align}
    \dot{a}_j=-(i\Delta\omega_j+\gamma/2)a_j-i\sum_{k}g_{jk}a_pb^\dag_{jk}+\eta_j,
    \label{eq:j_HL}
\end{align}
\begin{align}
    \dot{b}_{jk}=-(i\Delta\Omega_{jk}+\Gamma_{jk}/2)b_{jk}-ig_{jk}a_pa^\dag_{j}+\xi_{jk},
    \label{eq:phononHL}
\end{align}
where $\gamma$ is the optical decay rate and $\gamma_{ext}$ is the external coupling rate. The latter describes the coupling between the resonator and the bus waveguide, which supplies the external pump laser. The pump laser amplitude $F$ is normalized such that $|F|^2$ is given in units of photon flux, and relates to the externally supplied power by $P=\hbar\omega_p|F|^2$. Time dependent phases have been absorbed into the respective optical and acoustic Langevin forces, $\eta_{p/j}$ and $\xi_{jk}$, which describe the quantum and thermal fluctuations \cite{behunin_fundamental_2018, loh_noise_2015, li_characterization_2012}. These are assumed to be zero-mean, white noise variables which have the correlation properties,
\begin{align}
    \langle \eta_m^\dag(t)\eta_{m'}(t')\rangle=\delta_{m,m'}\gamma N_m\delta(t-t')
    \label{eq:langevin_corr_optics1}
\end{align}
\begin{align}
    \langle \eta_m(t)\eta_{m'}^\dag(t')\rangle=\delta_{m,m'}\gamma(N_m+1)\delta(t-t')
    \label{eq:langevin_corr_optics2}
\end{align}
\begin{align}
    \langle \xi_{jk}^\dag(t)\xi_{j'k'}(t')\rangle=\delta_{jj'}\delta_{kk'}\Gamma_{jk}n_{jk}\delta(t-t')
\end{align}
\begin{align}
    \langle \xi_{jk}(t)\xi_{j'k'}^\dag(t')\rangle=\delta_{jj'}\delta_{kk'}\Gamma_{jk}(n_{jk}+1)\delta(t-t').
\end{align}
where $N_m$ and $n_{jk}$ are the optical and acoustic thermal populations respectively (i.e., $N_m=[\exp\{\hbar\omega_m/k_BT\}-1]^{-1}$ and $n_{jk}=[\exp\{\hbar\Omega_{jk}/k_BT\}-1]^{-1}$), with $m$ as either pump $p$, or Stokes index $j$. The assumption that the Langevin forces are zero-mean, white noise variables is justified under the rotating wave approximation and the phonon linewidth being much narrower than the thermal distribution of the bath (i.e., $\hbar\Gamma\ll k_BT$) \cite{vahala_back-action_2008}.

\begin{table}
    \centering
    \caption{Large mode volume parameters (taken from Ref. \cite{liu_large_2025}).}
    \begin{ruledtabular}
    \begin{tabular}{llr}
        $\gamma$ & $(2\pi) 2.55$ MHz & optical decay rate \\
        $\gamma_{ext}$ & $(2\pi) 0.95$ MHz  & external optical loss rate\\
        $\omega_p$ & $(2\pi)195.3$ THz & pump mode frequency\\
        $L$ & 4.0 m & resonator length\\
        FSR & $48.1$ MHz & free spectral range\\
        $v_{p/j}$ & $2.0\times10^8$ m/s & optical group velocity\\
    \end{tabular}
    \end{ruledtabular}
    \label{table1}
\end{table}

\begin{table}
    \centering
    \caption{Large mode volume parameters derived from finite element simulations based off of the experimental design in Ref. \cite{liu_large_2025}. The parameters $g_{jk}$, $\Omega_{jk}$, $\Gamma_{jk}$ are insensitive across the cavity modes with potential to lase and thus are assumed to be constant with respect to $j$.}
    \begin{ruledtabular}
    \begin{tabular}{lccr}
        Mode ($k$) & $g_{jk}$ [Hz] & $\Omega_{jk}/(2\pi)$ [GHz] & $\Gamma_{jk}/(2\pi)$ [MHz] \\
        0 & 117.5 & 10.87 & 33\\
        1 & 94.5 & 10.93 & 30\\
        2 & 68.5 & 10.97 & 40\\
        3 & 78 & 11.07 & 40\\
        4 & 58.5 & 11.19 & 27\\
        5 & 41 & 11.27 & 36\\
        6 & 38 & 11.39 & 90\\
    \end{tabular}
    \end{ruledtabular}
    \label{table2}
\end{table}

\subsection{Adiabatic elimination}
Here, we solve the Heisenberg-Langevin equation for the phonon (Eq. \eqref{eq:phononHL}) to gather the time-dynamics of the acoustic fields. The solution to Eq. \eqref{eq:phononHL} is
\begin{align}
    b_{jk}(t)=\int_{-\infty}^t&d\tau e^{-(i\Delta\Omega_{jk}+\frac{\Gamma_{jk}}{2})(t-\tau)}\label{eq:integral}\\&\times\Bigl(\xi_{jk}(\tau)-ig_{jk}a_p(\tau)a_j^\dag(\tau)\Bigr).\nonumber
\end{align}
When the phonon decay rate greatly exceeds the optical decay rate ($\Gamma_{jk}\gg\gamma$), common for Brillouin lasers, the phonon field adiabatically follows the electrostrictive force produced by the optical beat note. In other words, the optical fields are assumed to vary slowly in time, so that the phonon field reaches its steady state amplitude at each instant. Under this condition, the integral in Eq. \eqref{eq:integral} can be approximated by writing $a_p(\tau)a_j^\dag(\tau)\simeq a_p(t)a_j^\dag(t)$, leading to the solution
\begin{align}
    b_{jk}\approx-ig_{jk}\chi_{jk}a_pa_j^\dag  + \hat{b}_{jk},
    \label{eq:ph_eq}
\end{align}
where $\chi_{jk}=(i\Delta\Omega_{jk}+\Gamma_{jk}/2)^{-1}$ and $\hat{b}_{jk}$ quantifies the thermal and quantum fluctuations of the phonon given by
\begin{align}
    \hat{b}_{jk}=\int_{-\infty}^td\tau e^{-(i\Delta\Omega_{jk}+\frac{\Gamma_{jk}}{2})(t-\tau)}\xi_{jk}(\tau),
    \label{eq:dressed_ph_L}
\end{align}
with two-time correlation function
\begin{align}
    \langle\hat{b}_{jk}^\dag (t) \hat{b}_{j'k'}(t')&\rangle\!=\!n_{th}\delta_{jj'}\delta_{kk'}
    \label{eq:bhatcorrelation}e^{-\frac{\Gamma_{jk}}{2}|t-t'|}e^{i\Delta\Omega_{jk}(t-t')}.
\end{align}
As experiments scale to longer resonators, a regime may be reached where treatment of this physics with adiabatic elimination is no longer appropriate, as it results in a divergent description of amplitude at a critical feedback parameter. This is overcome in a `three-wave model', which predicts pulsed steady state dynamics at the critical value \cite{montes_bifurcation_1994}.

Inserting Eq. \eqref{eq:ph_eq} into Eqs. \eqref{eq:p_HL} and \eqref{eq:j_HL}, we find coupled equations for the optical modes,
\begin{align}
    \dot{a}_p=-\bigl(i\Delta\omega_p&+\gamma/2+\sum_j\mu^*_ja^\dag_ja_j\bigr)a_p
    \label{eq:p_eq}\\&+\sqrt{\gamma_{ext}}F+h_p\nonumber
\end{align}
\begin{align}
    \dot{a}_j=-\bigl(i\Delta\omega_j+\gamma/2-\mu_ja_p^\dag a_p\bigr)a_j + h_j.
    \label{eq:j_eq}
\end{align}
Here, $\mu_j=\sum_k|g_{jk}|^2\chi_{jk}^*$, $h_p=\eta_p-i\sum_{jk}g_{jk}^*a_j\hat{b}_{jk}$, and $h_j=\eta_j-i\sum_{jk}g_{k}^*a_p\hat{b}_{jk}^\dag.$ The Langevin forces $h_{p/j}$ describe how  colored multiplicative noise is introduced on the optical fields via spontaneous Brillouin scattering, beyond the intrinsic quantum and thermal fluctuations of the optical modes. The function $\mu_j$ represents a nonlinear susceptibility associated with Brillouin scattering, where Re[$\mu_j$] is proportional to the Brillouin gain factor for each cavity mode, which is given by
\begin{align}
    G_{B,j}=\frac{2{\rm Re}[\mu_j]L}{\hbar\omega_j v_{p}v_j}.
\end{align}
The Brillouin gain factor is a constant that is determined by properties of the waveguide.

\subsection{Steady state dynamics}
\begin{figure}
    \centering
    \includegraphics[width=8.6cm]{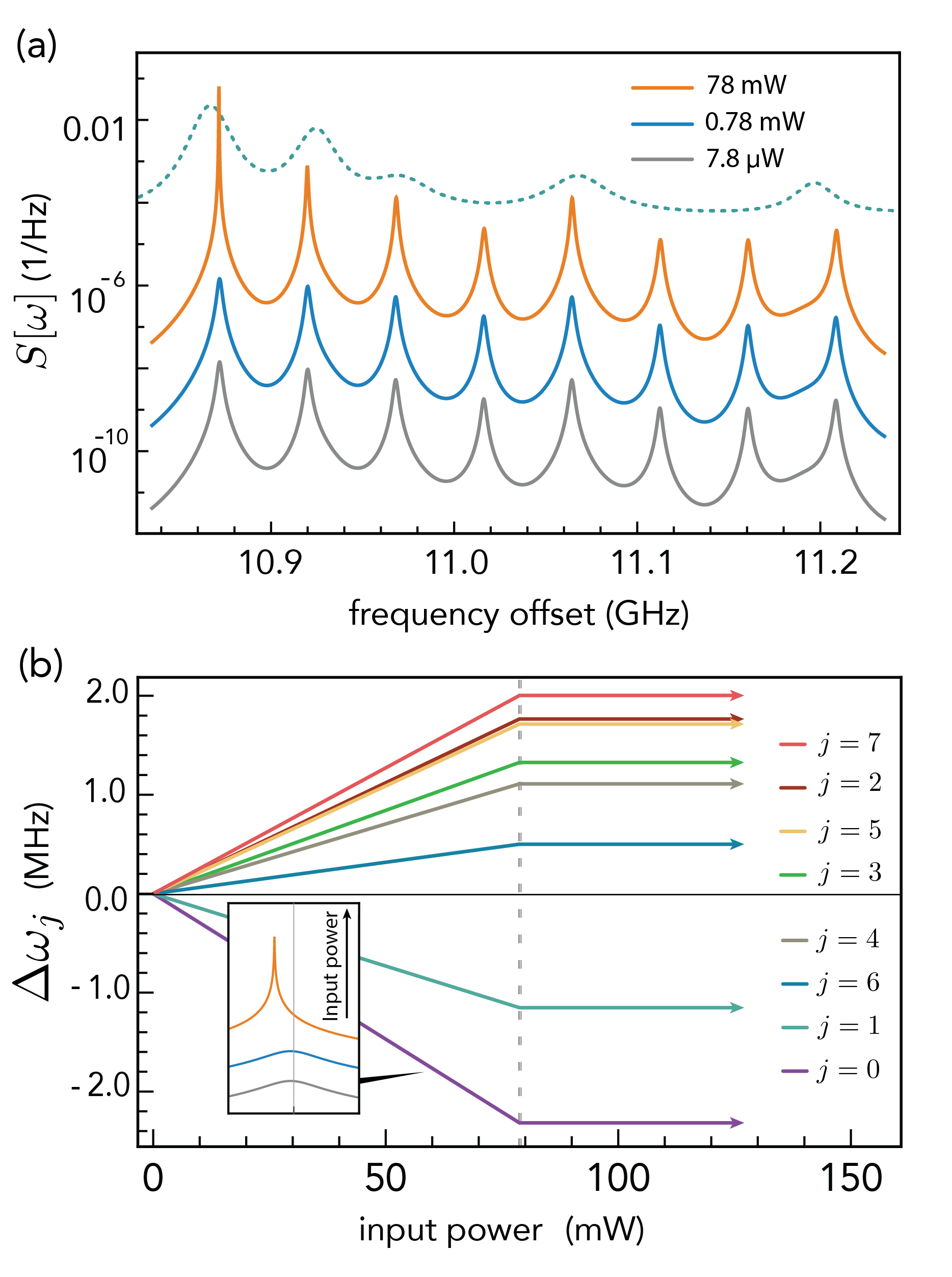}
    \caption{(a) Spontaneous power spectrum below threshold with increasing input power from gray to orange.  The gain spectrum shape is shown as the dotted line, with no relation to the y-axis. (b) Frequency pulling of each mode from cavity resonance vs. increasing input power. The vertical, dashed gray line represents $P^P_{th}$, where the frequency drift clamps with the pump. Inset: zoomed in power spectrum of the $j=0$ mode as power increases, where frequency pulling is contrasted with the cavity resonance, given by the gray vertical line.} 
    \label{fig:spontaneous}   
\end{figure}

We analyze the amplitudes of the optical modes in steady state (denoted here as $\alpha_p$ and $\alpha_j$) by dropping fluctuating terms and setting derivatives to zero Eqs. \eqref{eq:p_eq} and \eqref{eq:j_eq}, giving
\begin{align}
    \alpha_p(i\Delta\omega_p+\gamma/2+\sum_j\mu_j^*\alpha_j^2)=\sqrt{\gamma_{ext}}F
    \label{eq:ss_p}
\end{align}
\begin{align}
    \alpha_j(i\Delta\omega_j + \gamma/2 - \mu_j \alpha_p^2)=0
    \label{eq:ss_j}.
\end{align}
Equation \eqref{eq:ss_j} has two solutions: below threshold when all $\alpha_j=0$, and above threshold when $i\Delta\omega_j+\gamma/2-\mu_j\alpha_p^2=0$. The relation for above threshold is satisfied when both real and imaginary parts vanish independently. This results in two conditions. (1) The cavity mode with the highest gain, ${\rm Re}[\mu_l]>{\rm Re}[\mu_{j\neq l}]$, will satisfy the threshold relation first, giving the clamping condition for the pump,
\begin{align}
    \alpha_p^2=\frac{\gamma}{2\mu'_l},
    \label{eq:pump_clamp}
\end{align}
and will be the mode to lase, and (2) the frequency pulling of the lasing mode is 
\begin{align}
    \Delta\omega_l=\alpha_p^2\mu_l''=\frac{\gamma \mu_l''}{2\mu_l'}.
\end{align}
Here, we've introduced the notation $\mu'_j=\rm{Re}[\mu_j]$ and ${\rm Im}[\mu_l]=\mu''_l$. The frequency pulling is a consequence of phase matching not being satisfied between the beat note and the multiple phonon modes (i.e., $\mu_l''\neq 0$). This feature is somewhat unique as Brillouin lasers can be made resonant. All other modes remain with a steady state amplitude of zero, $\alpha_{j\neq l}=0$, as $(i\Delta\omega_{j\neq l}+\gamma/2-\mu_{j\neq l} \alpha_p^2)\neq 0$, showing that although multiple optical modes experience gain, only one mode lases. Once threshold is reached, the lasing cavity mode amplitude grows with the injected laser amplitude $F$, and is found using Eqs. \eqref{eq:ss_p} and \eqref{eq:pump_clamp},
\begin{align}
    \alpha_l^2=\frac{1}{\mu_l'}\Bigl(\frac{\sqrt{\gamma_{ext}}F}{\alpha_p}-\frac{\gamma}{2}\Bigr).
    \label{lasing_amp}
\end{align}
The pump frequency pulls with increasing lasing amplitude at $\Delta\omega_p=-\mu''_l\alpha_l^2$, shown in the inset of Fig \ref{fig:spontaneous}(b). 

Below threshold, there is no pump frequency pulling ($\Delta\omega_p=0$), and the pump amplitude growing with $F$ as 
\begin{align}
    \alpha_p=\frac{2\sqrt{\gamma_{ext}}F}{\gamma}.
    \label{eq:pump_before_lasing}
\end{align}
Combining Eqs. \eqref{eq:pump_clamp} and \eqref{eq:pump_before_lasing}, and using the relation between $F$ and input power $P$, we find expression for the pump threshold power,
\begin{align}
    P^P_{\rm th}=\frac{\gamma^3\hbar\omega_p}{8\mu_l'\gamma_{ext}}.
\end{align}
Using the values from Tables \ref{table1} and \ref{table2}, we find the first cavity mode has the highest gain (i.e., $\mu_l'=\mu_0'$) and $P^P_{th}=78.8$ mW.

The lasing cavity mode will climb in amplitude, with $\alpha_p$ given by the pump clamping condition in Eq. \eqref{eq:pump_clamp}, until it reaches threshold for cascaded order Brillouin lasing $\alpha_l^2=\gamma/(2\mu_{l2}')$, where $\mu_{l2}'$ relates to the gain of the cascaded lasing mode \cite{behunin_fundamental_2018}. Using this result, and assuming that $\mu_{l2}'\simeq\mu_l'$, we find the threshold of supplied power to reach cascaded lasing,
\begin{align}
    P^{S1}_{\rm th}=\frac{\gamma^3\hbar\omega_p}{8\mu_l'\gamma_{ext}}\Bigl(\frac{\mu_l'}{\mu_{l2}'}+1\Bigr)^2\simeq4P_{\rm th}^P.
    \label{eq:cascading}
\end{align}
Noting that output power is given by $P_{\rm out}=\hbar\omega_p\gamma_{ext}\alpha_l^2$, the output power at $S1$ cascading can be written as
\begin{align}
    P^{S1}_{\rm out}=\frac{\gamma_{ext}\gamma\hbar\omega_p}{2\mu_l'}=\frac{\gamma_{ext}\gamma L}{G_{B,l}v_p v_l},
\end{align}
evaluated as 43.7 mW. The external coupling can be described in terms of the length, coupling factor $\kappa$, and group velocity as $\gamma_{ext}=v_g\kappa/L$. By fixing $\gamma_{ext}$ (i.e., tailoring the bus-waveguide distance so that $\kappa/L$ remains constant with increasing $L$), lasing threshold powers $P_{\rm th}^P$ and $P^{S1}_{\rm th}$ grow linearly with increasing length. 

Standard Brillouin lasers can further increase output power by negating the onset of cascading with destructive interference via changes the resonator structure \cite{puckett_higher_2019, liu_integrated_2024}. In a large mode volume Brillouin laser, the altered structure would need to remove a broad band of frequencies, approximately equal to that of the phonon gain bandwidth. This difference is illustrated in Fig. \ref{fig:diagram}(c), where removal of the lasing mode from the cavity-supported modes does not negate cascading, as the mode with the next highest gain will lase.

\subsection{Spontaneous scattering}
Here, we solve for the spontaneous scattering power spectrum both before and after lasing occurs by solving the coupled Eqs. \eqref{eq:p_eq} and \eqref{eq:j_eq}. This is done using the steady state solution for $\alpha_p$ (either Eq. \eqref{eq:pump_before_lasing} before threshold or Eq. \eqref{eq:pump_clamp} above threshold), and taking the Fourier transform of Eq. \eqref{eq:j_eq}. This gives $a_j(\omega)=\ell_j(\omega)h_j(\omega)$ with $\ell_j(\omega)=(-i\omega+i\Delta\omega_j+\gamma/2-\mu_j\alpha_p^2)^{-1}$. Using this, the power spectrum for each mode $j$ is obtained by calculating
\begin{align}
    S_j[\omega]=\frac{\langle a^\dag_j(\omega)a_{j'}(\omega')\rangle}{2\pi\delta(\omega-\omega')}=|\ell_j(\omega)|^2S_{h_j}[\omega].
    \label{eq:powerspec}
\end{align}
The power spectrum for the Langevin force $h_j$ can be found using the stationary two time correlation function,
\begin{align}
    \langle h_j^{\dag} &(t) h_{j'} (t') \rangle \equiv \delta_{jj'}C_{h_j}(\tau)= \delta_{jj'}\Bigl(\gamma N_j\delta(\tau)\nonumber\\&+\sum_k|g_{jk}|^2 \alpha_p^2(n_{jk}+1)e^{-\frac{\Gamma_{jk}}{2}|\tau|}e^{-i\Delta\Omega_{jk}\tau}\Bigr),
    \label{hj_corr_spontaneous}
\end{align}
where $t-t'=\tau$, and the Wiener-Khinchine theorem,
\begin{align}
    S_{h_j}[\omega]=\int_{-\infty}^{\infty}d\tau e^{i\omega \tau} C_{h_j}(\tau).
    \label{eq:WK}
\end{align}
Inserting the solution of Eq. \eqref{eq:WK} into Eq. \eqref{eq:powerspec}, exiting the rotating frame, and discarding the negligible $\gamma N_j$, we find the full spontaneous power spectrum by summing over $j$,
\begin{align}
    S[\omega]=\sum_{jk}|\ell_j(\omega)|^2\frac{\alpha_p^2|g_{jk}|^2(n_{jk}+1)\Gamma_{jk}}{(\Gamma_{jk}/2)^2+(\omega+\omega'_p-\Omega_{jk})^2},
    \label{eq:spont_spec}
\end{align}  
noting that the cavity modes' individual power spectra are uncorrelated, and $\omega_j'-\Delta\Omega_{jk}=\omega_p'-\Omega_{jk}$.

Equation \eqref{eq:spont_spec} is plotted in Fig. \ref{fig:spontaneous}(a) for varying input powers below threshold, where $\alpha_p$ depends on the input power and $\omega_p'=\omega_p$. The gain shape is shown as the dotted line, to emphasize the influence of $\mu_j$ in the spontaneous scattering. Examining Eq. \eqref{eq:spont_spec} below threshold reveals that each mode experiences frequency pulling, given by $\Delta\omega_j=-\mu_j''\alpha_p^2$, and an effective narrowing of linewidth given by $\gamma_{{\rm eff},j}=\gamma-2\mu_j'\alpha_p^2$. As a consequence of these dynamics, the center frequency and linewidth of each cavity mode is fixed once the pump clamps at threshold. The frequency pulling is plotted against increasing power in Fig. \ref{fig:spontaneous}(b) for each mode, showing a  maximum drift magnitude ranging between $0.5-2.4$ MHz at $P_{\rm th}^P$ for each mode. Additionally, the magnitude of linewidth reduction ranges from $0.09-0.9$ MHz. With this information, paired with the steady state values that relate to the lasing mode, the real and imaginary parts of $\mu_j$ can be obtained for all cavity modes within the phonon gain bandwidth. This means that by simply reaching threshold in a large mode volume Brillouin laser, and measuring the linewidths and frequency drifts of each mode, full construction of the phonon gain spectrum is possible. By fitting a multiple-oscillator model to that spectrum, all values of $g_{jk}$, $\Omega_{jk}$, and $\Gamma_{jk}$ can be determined, making this system an interesting tool for phonon spectroscopy. 

The narrowed linewidth of the mode with the second highest gain could be the foundation for an interesting application in terms of a high effective $Q$ resonator mode. In the 4 meter resonator coil, the mode with the second highest gain sees an effective $1.5\times$ increase in $Q$ at $P_{\rm th}^P$, resulting in $Q_{eff}=\omega_p/\gamma_{eff}=194$ million, compared to the fabricated loaded quality factor of $77$ million. Although this is only a modest increase, longer resonators will bring the cavity modes closer together in frequency. In turn, the modes adjacent to the lasing mode will begin to approach equal gain, resulting in $\gamma_{{\rm eff},j}$ trending closer to zero, and increasing the effective quality factor. For example, when calculating the dynamics of this system with a resonator length of $50$ m, the quality factor sees a $75\times$ increase at $P_{\rm th}^P$, making $Q_{eff}=5.9$ billion. Examining Eq. \eqref{eq:spont_spec} above threshold, the spontaneous spectrum appears to be dependent on increasing input power as $\omega_p'=\omega_p+\mu_l''\alpha_l^2$, and $\alpha_l$ grows with input power; however, the frequency pulling of the pump is found to be effectively negligible in changing the spontaneous spectrum.

\subsection{Above threshold: Lasing}
In this section we explore the dynamics of amplitude and frequency fluctuations, using Eq. \eqref{eq:j_eq} for the lasing cavity mode ($j=l$). When examining above laser threshold, it is convenient to decompose each field as 
\begin{align}
    a=(\alpha+\delta\alpha)e^{i\varphi},
    \label{eq:decomp}
\end{align}
which includes the steady state amplitude $\alpha$, fluctuations in amplitude $\delta\alpha$ which describe the RIN, and fluctuations to the phase $\varphi$ which describe phase/frequency noise. Additionally, we capture the transferred amplitude and phase noise from the external pump with the decomposition $F=(F_0+\delta F)e^{i\varphi_f}$, where $\varphi_f$ is assumed to randomly fluctuate in time with a variance determined by the external laser's linewidth $\Delta\nu_f$. We assume that the fluctuations are small compared to the amplitude, (i.e., $|\delta\alpha|\ll\alpha$). Additionally, when the resonator linewidth is much larger than the external laser linewidth, ($\gamma\gg2\pi\Delta\nu_f$), one finds that the pump phase noise adiabatically follows the phase of the external source \cite{debut_linewidth_2000, behunin_fundamental_2018, dallyn_thermal_2022}, such that
\begin{align}
    \varphi_p\approx\varphi_f.
    \label{eq:phase_p_fluc}
\end{align}
After inserting the decomposition from Eq. \eqref{eq:decomp} into the time-dependent equations for the optical modes in the rotating frame (Eqs. \eqref{eq:p_eq} and \eqref{eq:j_eq}), keeping only terms linear in $\delta\alpha$ and $\varphi$, incorporating steady state definitions where appropriate, and using Eq. \eqref{eq:phase_p_fluc}, we take the real and imaginary parts to find the following equations
\begin{align}
    \dot{\delta\alpha}_p\approx-2\mu_l'&\alpha_l\alpha_p\delta\alpha_l-\sqrt{\gamma_{ext}}F_0\frac{\delta\alpha_p}{\alpha_p}\nonumber\\&+{\rm Re}[\tilde{h}_p]+\sqrt{\gamma_{ext}}\delta F.
    \label{eq:ampfluct_p_eq}
\end{align}
\begin{align}
    \dot{\delta\alpha}_l=2\mu_l'\alpha_p\alpha_l\delta\alpha_p+{\rm Re}[\tilde{h}_l]
    \label{eq:ampfluct_l_eq}
\end{align}
\begin{align}
    \dot{\varphi}_l=2\mu_l''\alpha_p\delta\alpha_p+\frac{1}{\alpha_l}{\rm Im}[\tilde{h}_l].
    \label{eq:phase_eq}
\end{align}
The previous considerations have altered the quantum/thermal fluctuation's definition, now precisely represented as $h_p=\eta_p-i\sum_{k}g^*_{lk}\alpha_le^{i\varphi_l}\hat{b}_{lk}$ and $h_l=\eta_l-\sum_kig_{lk}\alpha_pe^{i\varphi_p}\hat{b}^\dag_{lk}$, differing from prior theoretical descriptions which assume phase matched coupling to a single phonon mode \cite{behunin_fundamental_2018, dallyn_thermal_2022}. Additionally, $\tilde{h}_p=h_pe^{-i\varphi_p}$, $\tilde{h}_l=h_le^{-i\varphi_l}$. Here, one can observe that the lasing Stokes' phase fluctuations couple to the pump's amplitude fluctuations as a consequence of this system lacking phase matching for each phonon mode ($\mu_l''\neq 0$).

\begin{figure}
    \centering
    \includegraphics[width=8.6cm]{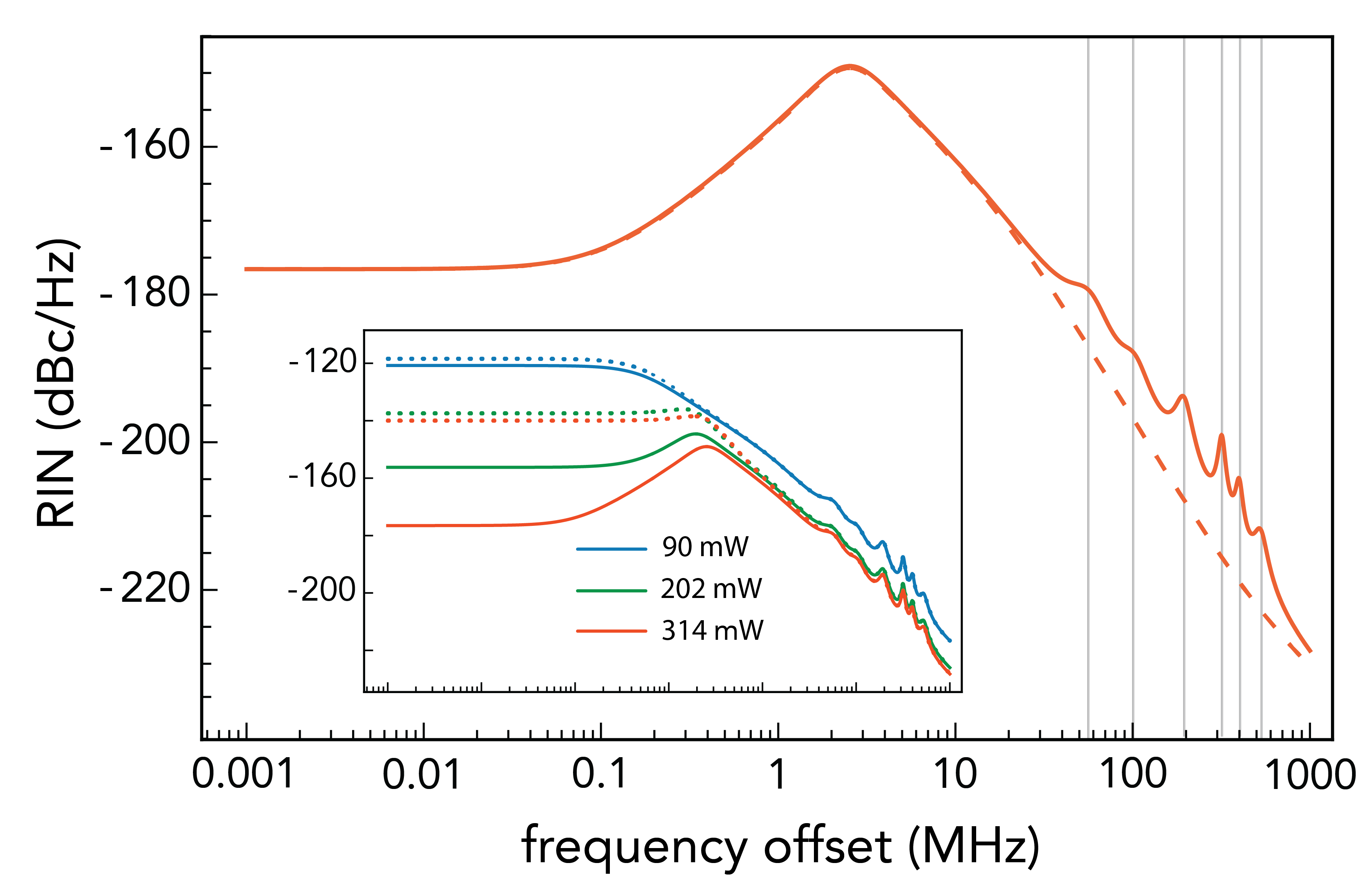}
    \caption{Theoretical RIN for the large mode volume Brillouin laser (solid red line) vs. a single mode laser involving overlap between one phonon and one cavity mode (dashed) with 314 mW of input power ($P_{th}^{S1}$). The vertical gray lines represent $\Delta\Omega_{lk}$'s, highlighting the source of additional noise at high-offset frequencies. Inset: Increasing input power from blue to red. Solid lines represent no transferred RIN, (i.e., $S_{ext}^{\rm RIN}=0$) and dotted lines show the inclusion of constant white noise, where $S_{ext}^{\rm RIN}=-140$ dBc/Hz.} 
    \label{fig:RIN}  
\end{figure}

\subsubsection{Relative intensity noise of lasing cavity mode}
In this section, we begin our analysis of the large mode volume laser noise by calculating the RIN, which is defined by the two-sided power spectrum
\begin{align}
    S^{{\rm RIN}}_l[\omega]=\frac{1}{P_l^2}\int_{-\infty}^{\infty}d\tau e^{i\omega\tau}\langle \delta P_l(t+\tau) \delta P_l(t) \rangle,
\end{align}
where the time-dependent variations of laser power are represented by $\delta P_l$. Noting that $(P_l+\delta P_l)\propto(\alpha_l+\delta\alpha_l)^2$ and $\delta\alpha_l$'s are small compared to the amplitude, we can express the RIN as
\begin{align}
    S^{{\rm RIN}}_l[\omega]=\frac{4}{\alpha_l^2}\int_{-\infty}^{\infty}d\tau e^{i\omega\tau}\langle \delta \alpha_l(t+\tau) \delta \alpha_l(t) \rangle,
\end{align}
The amplitude coupling in Eqs. \eqref{eq:ampfluct_p_eq} and \eqref{eq:ampfluct_l_eq} leads to relaxation oscillations, with resonant frequency $\omega_0^{{\rm rel}}=2\mu_l'\alpha_l\alpha_p$ and damping rate $\Gamma_{{\rm RIN}}=\sqrt{\gamma_{ext}}F_0/\alpha_p$. We solve these coupled equations in the Fourier domain for $\delta\alpha_l$, giving
\begin{align}
    \delta\alpha_l(\omega)=\chi_{{\rm RIN}}&(\omega)\bigl(\sqrt{\gamma_{ext}}\delta F(\omega)\omega_0^{{\rm rel}}+{\rm Re}[\tilde{h}_p(\omega)]\omega_0^{{\rm rel}}\nonumber\\&+(\Gamma_{{\rm RIN}}-i\omega){\rm Re}[\tilde{h}_l(\omega)] \bigr).
    \label{eq:FT_delta_alpha_l}
\end{align}
where $\chi_{{\rm RIN}}=(-\omega^2-i\omega\Gamma_{{\rm RIN}}+\omega_0^{{\rm rel^2}})^{-1}$. Using Eq. \eqref{eq:FT_delta_alpha_l} and the correlation properties of ${\rm Re}[\tilde{h}_p]$ and ${\rm Re}[\tilde{h}_l]$ (see Appendix A), we calculate the correlation function $\langle\delta\alpha_l^*(\omega)\delta\alpha_l(\omega')\rangle$. The resulting function is used to find the power spectrum for the RIN of a large mode volume Brillouin laser,
\begin{align}
    &S_l^{{\rm RIN}}[\omega]=\frac{|\chi_{{\rm RIN}}(\omega)|^2}{\alpha_l^2}\Biggl[ (\Gamma_{\rm RIN}\alpha_p\omega_0^{{\rm rel}})^2S^{\rm RIN}_{ext}[\omega]\label{eq:RIN}\\&+2\gamma(N_p+1/2)\omega_0^{{\rm rel}^2}+2\gamma(N_l+1/2)(\Gamma_{{\rm RIN}}^2+\omega^2)\nonumber\\&+\Bigl(\omega_0^{{\rm rel^2}}\alpha_l^2-2\omega_0^{{\rm rel}}\Gamma_{{\rm RIN}}\alpha_p\alpha_l+(\Gamma_{{\rm RIN}}^2+\omega^2)\alpha_p^2\Bigr)\beta_l(\omega)\Biggr],\nonumber
\end{align}
where,
\begin{align}
    \beta_l(\omega)=&\sum_k|g_{lk}|^2\biggl[\frac{(n_{lk}+1)\Gamma_{lk}}{\Gamma_{lk}^2/4+(\omega-\Delta\Omega_{lk})^2}\nonumber\\&+\frac{n_{lk}\Gamma_{lk}}{\Gamma_{lk}^2/4+(\omega+\Delta\Omega_{lk})^2}\biggr].
\end{align}
The transferred RIN from the external laser is given as $S^{\rm RIN}_{ext}[\omega]=4\langle \delta F(\omega)\delta F(\omega') \rangle/F_0^2$. 

Equation \eqref{eq:RIN} can be shown to reproduce the RIN of a single mode Brillouin laser \cite{loh_noise_2015, behunin_fundamental_2018} by assuming pump-Stokes coupling results from only one phase matched phonon mode (i.e., no sum on $k$ and $\Delta\Omega_l=0$, and assuming there is no transferred RIN). The RIN for the large mode volume Brillouin laser is plotted in Fig. \ref{fig:RIN} as a solid line, contrasted with the RIN of a standard Brillouin laser (dashed line), with an input power of $314$ mW ($P_{\rm th}^{S1}$). The added noise at high offset frequency can be attributed to $\beta(\omega)$, which differs as a consequence of coupling to multiple phonon modes lacking phase matching. This is highlighted by the gray vertical lines in Fig. \ref{fig:RIN}, which are placed at $\Delta\Omega_{lk}$ for $k=1$ to $6$. The inset of Fig. \ref{fig:RIN} shows the effect of increasing input power from just above $P_{\rm th}^P$ to $P_{\rm th}^{S1}$. The solid lines correspond to no transferred RIN from the external pump, and the dotted lines assume a constant white noise spectrum for the external RIN at $-140$ dBc/Hz.

\begin{figure}
    \centering
    \includegraphics[width=8.6cm]{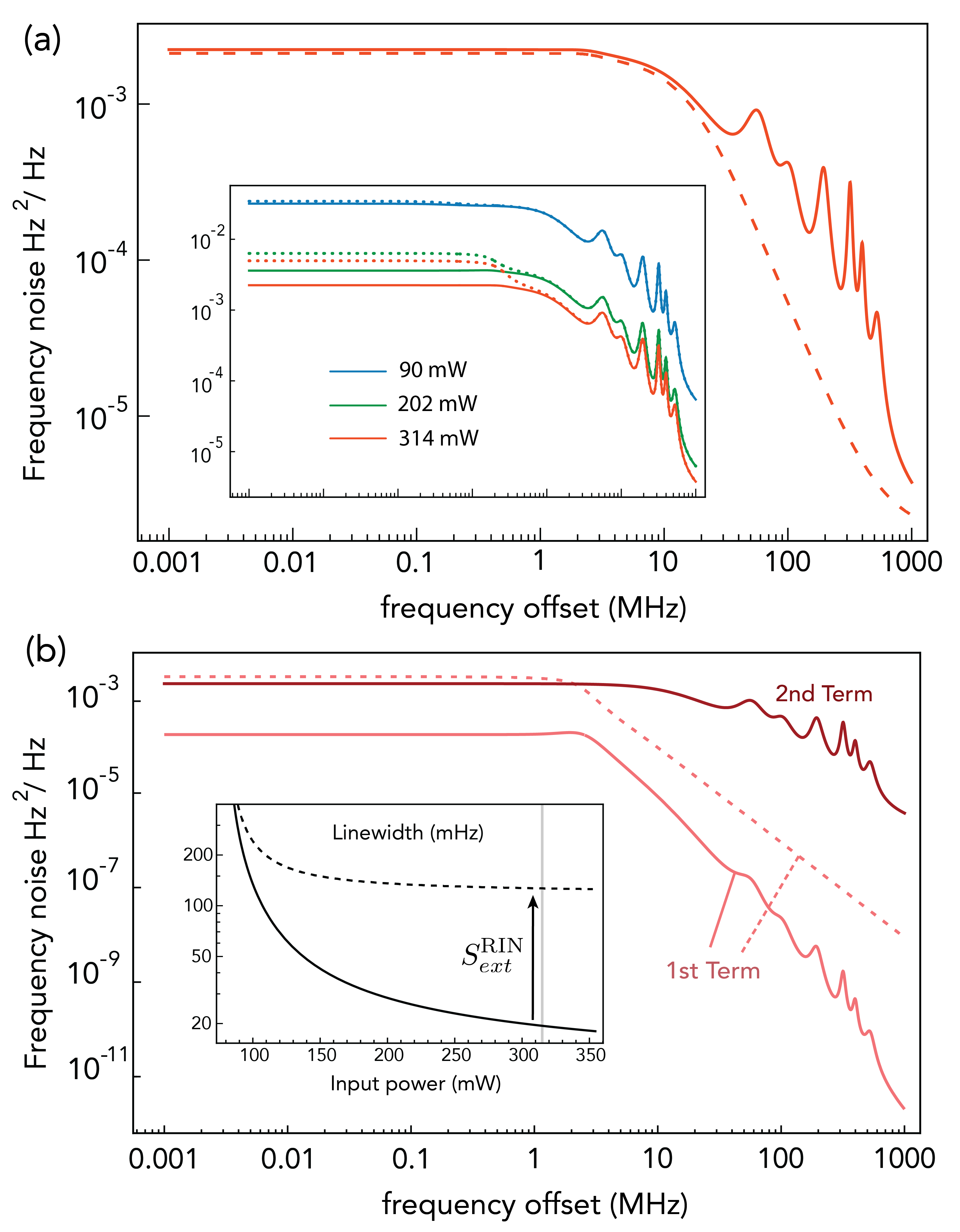}
    \caption{(a) Theoretical frequency noise spectrum for a large mode volume Brillouin laser (solid red line) vs. a single mode laser involving overlap between one phonon and one cavity mode (dashed) with an input power of 314 mW ($P_{th}^{S1}$). Inset: increasing input power from blue to red, and $S_{ext}^{\rm RIN}=0$ (solid) and $-140$ dBc/Hz (dotted). (b) Plotting $S_\Delta[\omega]$ and the second term of Eq. \eqref{eq:freqnoise} separated for $S_{ext}^{\rm RIN}=0$ (solid) and $-140$ dBc/Hz (dashed). Inset: Effect of $0\rightarrow-140$ dBc/Hz (solid $\rightarrow$ dashed) transferred external RIN on linewidth in the low-frequency limit (i.e., contents in the brackets of Eq. \eqref{eq:STLW}). The vertical line represents $P_{th}^{\rm S1}$.} 
    \label{fig:freqnoise}   
\end{figure}

\subsubsection{Frequency Noise of lasing cavity mode}
We now investigate the noise of the lasing mode frequency, $\nu_l$, where $2\pi\nu_l(t)=\dot{\varphi_l}$. As seen previously in Eq. \eqref{eq:phase_eq}, the frequency noise and amplitude noise of the pump are coupled due to the lack of phase matching between the optical beat note and the multiple phonon modes that make up the broad gain bandwidth ($\mu_l''\neq0$). In turn, this couples the lasing mode's phase and amplitude. The pump amplitude fluctuations in the Fourier domain can be expressed as
\begin{align}
    \delta\alpha_p(\omega)\!=\!\frac{-\omega_0^{{\rm rel}}\delta\alpha_l(\omega)+\sqrt{\gamma_{ext}}\delta F(\omega)+{\rm Re}[\tilde{h}_p(\omega)]}{-i\omega+\Gamma_{{\rm RIN}}},
\end{align}
yielding,
\begin{align}
    2\pi\nu_l(\omega)&=\frac{2\mu_l''\alpha_p}{-i\omega+\Gamma_{{\rm RIN}}}\Bigl(-\omega_0^{{\rm rel}}\delta\alpha_l(\omega)+\sqrt{\gamma_{ext}}\delta F(\omega)\nonumber\\&+{\rm Re}[\tilde{h}_p(\omega)]\Bigr)+\frac{1}{\alpha_l}{\rm Im}[\tilde{h}_l(\omega)].
    \label{eq:freq_variation}
\end{align}
Using Eq. \eqref{eq:freq_variation} and the correlation properties of $\tilde{h}_p$ and $\tilde{h}_l$ detailed in the Appendix A, we calculate $\langle\nu_l^*(\omega)\nu_l(\omega')\rangle$, which provides the two-sided power spectrum for the frequency noise as
\begin{align}
     S_{\nu_l}[\omega]\!=\!S_{\Delta}[\omega]\!+\!\frac{1}{8\pi^2\alpha_l^2}\Bigl(\gamma(N_l\!+\!1/2)+\frac{1}{2}\alpha_p^2\beta_l(\omega)\Bigr),
     \label{eq:freqnoise}
\end{align}
where $S_{\Delta}[\omega]$ is a consequence of phase-amplitude coupling, given by
\begin{align}
    S_{\Delta}[\omega]=\frac{1}{4\pi^2}&\frac{\mu''^2_l\alpha_p^2}{\omega^2+\Gamma_{{\rm RIN}}^2}\Bigl(\alpha_l^2\omega_0^{{\rm rel}^2} S^{{\rm RIN}}_l[\omega]+2\gamma(N_p+1/2)\nonumber\\&+\Gamma_{\rm RIN}^2\alpha_p^2 S^{\rm RIN}_{ext}[\omega]+\alpha_l^2 \beta_l(\omega)\Bigr).
\end{align}
Here, both the RIN of the Brillouin laser and the RIN of the external pump have an effect on the frequency noise. The frequency noise spectra is shown in Fig. \ref{fig:freqnoise}(a) as a solid red line, contrasted with a single mode, phase matched Brillouin laser as the dashed red line. This contrast, arising from the lack of phase matching, appears more significant in the frequency noise compared to the RIN spectra. The inset of Fig. \ref{fig:freqnoise}(a) shows increasing power from blue to red, with no transferred external RIN (solid lines) vs. an assumed constant white noise external RIN of $S_{ext}^{\rm RIN}[\omega]=-140$ dBc/Hz (dashed), with the same values for power as Fig. \ref{fig:RIN}(b).

The contributions from $S_\Delta[\omega]$ and the second term of Eq. \eqref{eq:freqnoise} are shown separated in Fig \ref{fig:freqnoise}(b), where input power matches the solid red line in Fig. \ref{fig:freqnoise}(a). Here, assuming an ideal external laser (no RIN) leaves $S_\Delta$ an order of magnitude smaller (solid lines); however, with a modest transferred RIN of $-140$ dBc/Hz (dashed line), these two terms are of the same order. We analyze the impact of this on the linewidth by examining the low frequency limit,
\begin{align}
    S_{\nu_l}[\omega\sim0]\simeq &\nonumber\frac{1}{2\pi}\biggl[2\pi S_\Delta[\omega\sim0]\\&+\frac{\gamma}{4\pi\alpha_l^2}(N_l+n_{th}+1)\biggr],
    \label{eq:STLW}
\end{align}
where the quantity in the brackets is the modified Schawlow-Townes-like linewidth for this system when an ideal external pump is assumed, with the low-offset frequency contribution from the inherent phase-mismatch given by
\begin{align}
    S_\Delta[\omega\!\sim\!0]\!&\simeq\!\frac{\mu_l''^2\alpha_p^2}{4\pi^2\Gamma_{\rm RIN}^2}\biggl[4\gamma(N_p\!+\!1/2)\!+\!2\gamma\frac{\Gamma_{\rm RIN}^2}{\omega_0^{\rm rel^2}}(N_l\!+\!1/2)\nonumber\\&+2\Gamma_{\rm RIN}^2\alpha_p^2S_{ext}^{\rm RIN}[\omega\!\sim\!0]
    +8\mu_l'\biggl(\alpha_l^2
    \!\nonumber\\&-\frac{\alpha_l\Gamma_{\rm RIN}\alpha_p}{\omega_0^{\rm rel}}+\!\frac{\Gamma_{\rm RIN}^2\alpha_p^2}{2\omega_0^{\rm rel^2}}\biggr)(n_{th}\!+\!1/2)\biggr].
    \label{eq:s_delta_STLW}
\end{align}
The inset of Fig. \ref{fig:freqnoise}(b) shows the significant implications of external RIN in Brillouin lasers that lack ideal phase matching, where the fundamental linewidth (Eq. \eqref{eq:STLW}) is plotted vs increasing input power for $S_{ext}^{\rm RIN}=0$ (solid line) and $S_{ext}^{\rm RIN}=-140$ dBc/Hz (dashed line). Although this sensitivity is inherent to large mode volume Brillouin lasers, this result emphasizes the importance of operating traditional Brillouin lasers in perfect phase matching conditions for lowest achievable linewidths.

\section{conclusion}
In this paper, we have developed a multiple oscillator model to understand the dynamics and noise properties  of large mode volume Brillouin lasers. Through a coupled-mode approach, we model the unique dynamics of a large mode volume system where multiple cavity modes have the potential to lase given a small FSR and broad Brillouin gain bandwidth. Using experimentally determined parameters in our model, we capture the key features of spontaneous scattering, steady-state dynamics, amplitude fluctuations, and frequency fluctuations. Notably, the spontaneous calculation reveals that the frequency pulling and linewidth narrowing of the cavity modes below lasing threshold may be used to uncover all salient features of the phonon field, and the non-lasing modes above threshold may result in ultra-high effective quality factors as designs scale to longer resonators.

The steady-state dynamics confirm the onset of single-mode lasing, despite the many optical modes lying within the gain bandwidth. Through characterization of amplitude fluctuations, we find a RIN spectra that essentially reproduces that of a single-mode Brillouin laser; however, the intrinsic off-resonant coupling to multiple phonon modes gives rise to unique noise features at frequency offsets equal to the degree of frequency mismatch, which are slightly higher than that of a standard Brillouin laser. We find this atypical noise feature is even more significant in the frequency noise spectrum. With inclusion of transferred external RIN in our analysis, we find that Brillouin lasers lacking phase matching between the optical beat note and the phonon mode--or modes--are sensitive to a noisy pump. Consequently, this may have a significant impact on the achievable linewidth.

{\it Acknowledgments}
This work was supported by NSF Awards No. 2145724 and No. 2427169. The authors thank Kaikai Liu for stimulating discussions.

\bibliography{references}

@article{vahala_back-action_2008,
	title = {Back-action limit of linewidth in an optomechanical oscillator},
	volume = {78},
	doi = {10.1103/PhysRevA.78.023832},
	number = {2},
	journal = {Physical Review A},
	author = {Vahala, Kerry J.},
	year = {2008},
}

@article{montes_bifurcation_1994,
	title = {Bifurcation in a cw-pumped {Brillouin} fiber-ring laser: {Coherent} soliton morphogenesis},
	volume = {49},
	shorttitle = {Bifurcation in a cw-pumped {Brillouin} fiber-ring laser},
	url = {https://link.aps.org/doi/10.1103/PhysRevA.49.1344},
	doi = {10.1103/PhysRevA.49.1344},
	number = {2},
	urldate = {2025-09-02},
	journal = {Physical Review A},
	author = {Montes, Carlos and Mamhoud, Abdellatif and Picholle, Eric},
	month = feb,
	year = {1994},
	note = {Publisher: American Physical Society},
	pages = {1344--1349},
}

@article{dallyn_thermal_2022,
	title = {Thermal and driven noise in {Brillouin} lasers},
	volume = {105},
	url = {https://link.aps.org/doi/10.1103/PhysRevA.105.043506},
	doi = {10.1103/PhysRevA.105.043506},
	number = {4},
	urldate = {2025-07-29},
	journal = {Physical Review A},
	author = {Dallyn, John H. and Liu, Kaikai and Harrington, Mark W. and Brodnik, Grant M. and Rakich, Peter T. and Blumenthal, Daniel J. and Behunin, Ryan O.},
	month = apr,
	year = {2022},
	note = {Publisher: American Physical Society},
	pages = {043506},
}

@article{debut_linewidth_2000,
	title = {Linewidth narrowing in {Brillouin} lasers: {Theoretical} analysis},
	volume = {62},
	shorttitle = {Linewidth narrowing in {Brillouin} lasers},
	url = {https://link.aps.org/doi/10.1103/PhysRevA.62.023803},
	doi = {10.1103/PhysRevA.62.023803},
	number = {2},
	urldate = {2025-07-28},
	journal = {Physical Review A},
	author = {Debut, Alexis and Randoux, Stéphane and Zemmouri, Jaouad},
	month = jul,
	year = {2000},
	note = {Publisher: American Physical Society},
	pages = {023803},
}

@article{loh_noise_2015,
	title = {Noise and dynamics of stimulated-{Brillouin}-scattering microresonator lasers},
	volume = {91},
	url = {https://link.aps.org/doi/10.1103/PhysRevA.91.053843},
	doi = {10.1103/PhysRevA.91.053843},
	number = {5},
	urldate = {2025-07-28},
	journal = {Physical Review A},
	author = {Loh, William and Papp, Scott B. and Diddams, Scott A.},
	month = may,
	year = {2015},
	note = {Publisher: American Physical Society},
	pages = {053843},
}

@article{huang_thermorefractive_2019,
	title = {Thermorefractive noise in silicon-nitride microresonators},
	volume = {99},
	url = {https://link.aps.org/doi/10.1103/PhysRevA.99.061801},
	doi = {10.1103/PhysRevA.99.061801},
	number = {6},
	urldate = {2025-07-25},
	journal = {Physical Review A},
	author = {Huang, Guanhao and Lucas, Erwan and Liu, Junqiu and Raja, Arslan S. and Lihachev, Grigory and Gorodetsky, Michael L. and Engelsen, Nils J. and Kippenberg, Tobias J.},
	month = jun,
	year = {2019},
	note = {Publisher: American Physical Society},
	pages = {061801},
}

@article{gorodetsky_fundamental_2004,
	title = {Fundamental thermal fluctuations in microspheres},
	copyright = {© 2004 Optical Society of America},
	url = {https://opg.optica.org/josab/abstract.cfm?uri=josab-21-4-697},
	doi = {10.1364/JOSAB.21.000697},
	language = {EN},
	urldate = {2025-07-25},
	journal = {JOSA B, Vol. 21, Issue 4, pp. 697-705},
	author = {Gorodetsky, Michael L. and Grudinin, Ivan S.},
	month = apr,
	year = {2004},
	note = {Publisher: Optica Publishing Group},
}

@article{qin_high-power_2022,
	title = {High-power, low-noise {Brillouin} laser on a silicon chip},
	volume = {47},
	copyright = {© 2022 Optica Publishing Group},
	issn = {1539-4794},
	url = {https://opg.optica.org/ol/abstract.cfm?uri=ol-47-7-1638},
	doi = {10.1364/OL.455369},
	language = {EN},
	number = {7},
	urldate = {2025-07-25},
	journal = {Optics Letters},
	author = {Qin, Yingchun and Ding, Shulin and Zhang, Menghua and Wang, Yunan and Shi, Qi and Li, Zhixuan and Wen, Jianming and Xiao, Min and Jiang, Xiaoshun},
	month = apr,
	year = {2022},
	note = {Publisher: Optica Publishing Group},
	pages = {1638--1641},
}

@article{li_characterization_2012,
	title = {Characterization of a high coherence, {Brillouin} microcavity laser on silicon},
	volume = {20},
	copyright = {© 2012 OSA},
	issn = {1094-4087},
	url = {https://opg.optica.org/oe/abstract.cfm?uri=oe-20-18-20170},
	doi = {10.1364/OE.20.020170},
	language = {EN},
	number = {18},
	urldate = {2025-07-25},
	journal = {Optics Express},
	author = {Li, Jiang and Lee, Hansuek and Chen, Tong and Vahala, Kerry J.},
	month = aug,
	year = {2012},
	note = {Publisher: Optica Publishing Group},
	pages = {20170--20180},
}

@article{lee_chemically_2012,
	title = {Chemically etched ultrahigh-{Q} wedge-resonator on a silicon chip},
	volume = {6},
	copyright = {2012 Springer Nature Limited},
	issn = {1749-4893},
	url = {https://www.nature.com/articles/nphoton.2012.109},
	doi = {10.1038/nphoton.2012.109},
	language = {en},
	number = {6},
	urldate = {2025-07-25},
	journal = {Nature Photonics},
	author = {Lee, Hansuek and Chen, Tong and Li, Jiang and Yang, Ki Youl and Jeon, Seokmin and Painter, Oskar and Vahala, Kerry J.},
	month = jun,
	year = {2012},
	note = {Publisher: Nature Publishing Group},
	pages = {369--373},
}

@article{lu_emerging_2024,
	title = {Emerging integrated laser technologies in the visible and short near-infrared regimes},
	volume = {18},
	copyright = {2024 This is a U.S. Government work and not under copyright protection in the US; foreign copyright protection may apply},
	issn = {1749-4893},
	url = {https://www.nature.com/articles/s41566-024-01529-5},
	doi = {10.1038/s41566-024-01529-5},
	language = {en},
	number = {10},
	urldate = {2025-07-25},
	journal = {Nature Photonics},
	author = {Lu, Xiyuan and Chang, Lin and Tran, Minh A. and Komljenovic, Tin and Bowers, John E. and Srinivasan, Kartik},
	month = oct,
	year = {2024},
	note = {Publisher: Nature Publishing Group},
	pages = {1010--1023},
}

@article{geng_pump--stokes_2007,
	title = {Pump-to-{Stokes} transfer of relative intensity noise in {Brillouin} fiber ring lasers},
	volume = {32},
	copyright = {© 2006 Optical Society of America},
	issn = {1539-4794},
	url = {https://opg.optica.org/ol/abstract.cfm?uri=ol-32-1-11},
	doi = {10.1364/OL.32.000011},
	language = {EN},
	number = {1},
	urldate = {2025-07-25},
	journal = {Optics Letters},
	author = {Geng, Jihong and Jiang, Shibin},
	month = jan,
	year = {2007},
	note = {Publisher: Optica Publishing Group},
	pages = {11--13},
}

@article{molin_experimental_2008,
	title = {Experimental investigation of relative intensity noise in {Brillouin} fiber ring lasers for microwave photonics applications},
	volume = {33},
	copyright = {© 2008 Optical Society of America},
	issn = {1539-4794},
	url = {https://opg.optica.org/ol/abstract.cfm?uri=ol-33-15-1681},
	doi = {10.1364/OL.33.001681},
	language = {EN},
	number = {15},
	urldate = {2025-07-25},
	journal = {Optics Letters},
	author = {Molin, Stéphanie and Baili, Ghaya and Alouini, Mehdi and Dolfi, Daniel and Huignard, Jean-Pierre},
	month = aug,
	year = {2008},
	note = {Publisher: Optica Publishing Group},
	pages = {1681--1683},
}

@article{stepien_intensity_2002,
	title = {Intensity noise in {Brillouin} fiber ring lasers},
	volume = {19},
	copyright = {© 2002 Optical Society of America},
	issn = {1520-8540},
	url = {https://opg.optica.org/josab/abstract.cfm?uri=josab-19-5-1055},
	doi = {10.1364/JOSAB.19.001055},
	language = {EN},
	number = {5},
	urldate = {2025-07-25},
	journal = {JOSA B},
	author = {Stépien, L. and Randoux, S. and Zemmouri, J.},
	month = may,
	year = {2002},
	note = {Publisher: Optica Publishing Group},
	pages = {1055--1066},
}

@article{smith_narrow-linewidth_1991,
	title = {Narrow-linewidth stimulated {Brillouin} fiber laser and applications},
	copyright = {© 1991 Optical Society of America},
	url = {https://opg.optica.org/ol/abstract.cfm?uri=ol-16-6-393},
	doi = {10.1364/OL.16.000393},
	language = {EN},
	urldate = {2025-07-25},
	journal = {Optics Letters, Vol. 16, Issue 6, pp. 393-395},
	author = {Smith, S. P. and Zarinetchi, F. and Ezekiel, S.},
	month = mar,
	year = {1991},
	note = {Publisher: Optica Publishing Group},
}

@article{debut_experimental_2001,
	title = {Experimental and theoretical study of linewidth narrowing in {Brillouin} fiber ring lasers},
	volume = {18},
	copyright = {© 2001 Optical Society of America},
	issn = {1520-8540},
	url = {https://opg.optica.org/josab/abstract.cfm?uri=josab-18-4-556},
	doi = {10.1364/JOSAB.18.000556},
	language = {EN},
	number = {4},
	urldate = {2025-07-25},
	journal = {JOSA B},
	author = {Debut, Alexis and Randoux, Stéphane and Zemmouri, Jaouad},
	month = apr,
	year = {2001},
	note = {Publisher: Optica Publishing Group},
	pages = {556--567},
}

@article{li_reaching_2021,
	title = {Reaching fiber-laser coherence in integrated photonics},
	volume = {46},
	copyright = {© 2021 Optical Society of America},
	issn = {1539-4794},
	url = {https://opg.optica.org/ol/abstract.cfm?uri=ol-46-20-5201},
	doi = {10.1364/OL.439720},
	language = {EN},
	number = {20},
	urldate = {2025-07-24},
	journal = {Optics Letters},
	author = {Li, Bohan and Jin, Warren and Wu, Lue and Chang, Lin and Wang, Heming and Shen, Boqiang and Yuan, Zhiquan and Feshali, Avi and Paniccia, Mario and Vahala, Kerry J. and Bowers, John E.},
	month = oct,
	year = {2021},
	note = {Publisher: Optica Publishing Group},
	pages = {5201--5204},
}

@article{isichenko_sub-hz_2024,
	title = {Sub-{Hz} fundamental, sub-{kHz} integral linewidth self-injection locked 780 nm hybrid integrated laser},
	volume = {14},
	copyright = {2024 The Author(s)},
	issn = {2045-2322},
	url = {https://www.nature.com/articles/s41598-024-76699-x},
	doi = {10.1038/s41598-024-76699-x},
	language = {en},
	number = {1},
	urldate = {2025-07-24},
	journal = {Scientific Reports},
	author = {Isichenko, Andrei and Hunter, Andrew S. and Bose, Debapam and Chauhan, Nitesh and Song, Meiting and Liu, Kaikai and Harrington, Mark W. and Blumenthal, Daniel J.},
	month = nov,
	year = {2024},
	note = {Publisher: Nature Publishing Group},
	pages = {27015},
}

@article{krinner_low_2024,
	title = {Low phase noise cavity transmission self-injection locked diode laser system for atomic physics experiments},
	volume = {32},
	issn = {1094-4087},
	url = {https://opg.optica.org/oe/abstract.cfm?uri=oe-32-9-15912},
	doi = {10.1364/OE.514247},
	language = {EN},
	number = {9},
	urldate = {2025-07-24},
	journal = {Optics Express},
	author = {Krinner, L. and Dietze, K. and Pelzer, L. and Spethmann, N. and Schmidt, P. O.},
	month = apr,
	year = {2024},
	note = {Publisher: Optica Publishing Group},
	pages = {15912--15922},
}

@article{matei_15_2017,
	title = {1.5 μ m {Lasers} with {Sub}-10 {mHz} {Linewidth}},
	volume = {118},
	copyright = {http://link.aps.org/licenses/aps-default-license},
	issn = {0031-9007, 1079-7114},
	url = {http://link.aps.org/doi/10.1103/PhysRevLett.118.263202},
	doi = {10.1103/PhysRevLett.118.263202},
	language = {en},
	number = {26},
	urldate = {2025-07-24},
	journal = {Physical Review Letters},
	author = {Matei, D. G. and Legero, T. and Häfner, S. and Grebing, C. and Weyrich, R. and Zhang, W. and Sonderhouse, L. and Robinson, J. M. and Ye, J. and Riehle, F. and Sterr, U.},
	month = jun,
	year = {2017},
	pages = {263202},
}

@article{liu_36_2022,
	title = {36 {Hz} integral linewidth laser based on a photonic integrated 4.0 m coil resonator},
	volume = {9},
	copyright = {© 2022 Optica Publishing Group},
	issn = {2334-2536},
	url = {https://opg.optica.org/optica/abstract.cfm?uri=optica-9-7-770},
	doi = {10.1364/OPTICA.451635},
	language = {EN},
	number = {7},
	urldate = {2025-07-24},
	journal = {Optica},
	author = {Liu, Kaikai and Chauhan, Nitesh and Wang, Jiawei and Isichenko, Andrei and Brodnik, Grant M. and Morton, Paul A. and Behunin, Ryan O. and Papp, Scott B. and Blumenthal, Daniel J.},
	month = jul,
	year = {2022},
	note = {Publisher: Optica Publishing Group},
	pages = {770--775},
}

@article{bai_comprehensive_2022,
	title = {A comprehensive review on the development and applications of narrow-linewidth lasers},
	volume = {64},
	issn = {1098-2760},
	url = {https://onlinelibrary.wiley.com/doi/abs/10.1002/mop.33046},
	doi = {10.1002/mop.33046},
	language = {en},
	number = {12},
	urldate = {2025-07-24},
	journal = {Microwave and Optical Technology Letters},
	author = {Bai, Zhenxu and Zhao, Zhongan and Tian, Menghan and Jin, Duo and Pang, Yajun and Li, Sensen and Yan, Xiusheng and Wang, Yulei and Lu, Zhiwei},
	year = {2022},
	note = {\_eprint: https://onlinelibrary.wiley.com/doi/pdf/10.1002/mop.33046},
	pages = {2244--2255},
}

@article{gundavarapu_sub-hertz_2019,
	title = {Sub-hertz fundamental linewidth photonic integrated {Brillouin} laser},
	volume = {13},
	copyright = {2018 The Author(s), under exclusive licence to Springer Nature Limited},
	issn = {1749-4893},
	url = {https://www.nature.com/articles/s41566-018-0313-2},
	doi = {10.1038/s41566-018-0313-2},
	language = {en},
	number = {1},
	urldate = {2025-07-24},
	journal = {Nature Photonics},
	author = {Gundavarapu, Sarat and Brodnik, Grant M. and Puckett, Matthew and Huffman, Taran and Bose, Debapam and Behunin, Ryan and Wu, Jianfeng and Qiu, Tiequn and Pinho, Cátia and Chauhan, Nitesh and Nohava, Jim and Rakich, Peter T. and Nelson, Karl D. and Salit, Mary and Blumenthal, Daniel J.},
	month = jan,
	year = {2019},
	note = {Publisher: Nature Publishing Group},
	pages = {60--67},
}

@article{pahlavani_linewidth_2025,
	title = {Linewidth narrowing in {Raman} lasers},
	volume = {10},
	issn = {2378-0967},
	url = {https://doi.org/10.1063/5.0271652},
	doi = {10.1063/5.0271652},
	number = {7},
	urldate = {2025-07-24},
	journal = {APL Photonics},
	author = {Pahlavani, R. L. and Spence, D. J. and Sharp, A. O. and Mildren, R. P.},
	month = jul,
	year = {2025},
	pages = {076107},
}

@article{heim_hybrid_2025,
	title = {Hybrid integrated ultra-low linewidth coil stabilized isolator-free widely tunable external cavity laser},
	volume = {16},
	copyright = {2025 The Author(s)},
	issn = {2041-1723},
	url = {https://www.nature.com/articles/s41467-025-61122-4},
	doi = {10.1038/s41467-025-61122-4},
	language = {en},
	number = {1},
	urldate = {2025-07-24},
	journal = {Nature Communications},
	author = {Heim, David A. S. and Bose, Debapam and Liu, Kaikai and Isichenko, Andrei and Blumenthal, Daniel J.},
	month = jul,
	year = {2025},
	note = {Publisher: Nature Publishing Group},
	pages = {5944},
}

@article{morton_high-power_2018,
	title = {High-{Power}, {Ultra}-{Low} {Noise} {Hybrid} {Lasers} for {Microwave} {Photonics} and {Optical} {Sensing}},
	volume = {36},
	issn = {1558-2213},
	url = {https://ieeexplore.ieee.org/document/8319492},
	doi = {10.1109/JLT.2018.2817175},
	number = {21},
	urldate = {2025-07-24},
	journal = {Journal of Lightwave Technology},
	author = {Morton, Paul A. and Morton, Michael J.},
	month = nov,
	year = {2018},
	pages = {5048--5057},
}

@article{wu_hybrid_2024,
	title = {Hybrid integrated tunable external cavity laser with sub-10 {Hz} intrinsic linewidth},
	volume = {9},
	issn = {2378-0967},
	url = {https://doi.org/10.1063/5.0190696},
	doi = {10.1063/5.0190696},
	number = {2},
	urldate = {2025-07-24},
	journal = {APL Photonics},
	author = {Wu, Yilin and Shao, Shuai and Tang, Liwei and Yang, Sigang and Chen, Hongwei and Chen, Minghua},
	month = feb,
	year = {2024},
	pages = {021302},
}

@article{fan_hybrid_2020,
	title = {Hybrid integrated {InP}-{Si}$_{\textrm{3}}${N}$_{\textrm{4}}$ diode laser with a 40-{Hz} intrinsic linewidth},
	volume = {28},
	copyright = {© 2020 Optical Society of America},
	issn = {1094-4087},
	url = {https://opg.optica.org/oe/abstract.cfm?uri=oe-28-15-21713},
	doi = {10.1364/OE.398906},
	language = {EN},
	number = {15},
	urldate = {2025-07-24},
	journal = {Optics Express},
	author = {Fan, Youwen and Rees, Albert van and Slot, Peter J. M. van der and Mak, Jesse and Oldenbeuving, Ruud M. and Hoekman, Marcel and Geskus, Dimitri and Roeloffzen, Chris G. H. and Boller, Klaus-J.},
	month = jul,
	year = {2020},
	note = {Publisher: Optica Publishing Group},
	pages = {21713--21728},
}

@article{li_microresonator_2017,
	title = {Microresonator {Brillouin} gyroscope},
	volume = {4},
	copyright = {© 2017 Optical Society of America},
	issn = {2334-2536},
	url = {https://opg.optica.org/optica/abstract.cfm?uri=optica-4-3-346},
	doi = {10.1364/OPTICA.4.000346},
	language = {EN},
	number = {3},
	urldate = {2025-07-24},
	journal = {Optica},
	author = {Li, Jiang and Suh, Myoung-Gyun and Vahala, Kerry},
	month = mar,
	year = {2017},
	note = {Publisher: Optica Publishing Group},
	pages = {346--348},
}

@article{liu_large_2025,
	title = {Large mode volume integrated {Brillouin} lasers for scalable ultra-low linewidth and high power},
	volume = {16},
	copyright = {2025 The Author(s)},
	issn = {2041-1723},
	url = {https://www.nature.com/articles/s41467-025-61637-w},
	doi = {10.1038/s41467-025-61637-w},
	language = {en},
	number = {1},
	urldate = {2025-07-23},
	journal = {Nature Communications},
	author = {Liu, Kaikai and Nelson, Karl D. and Behunin, Ryan O. and Blumenthal, Daniel J.},
	month = jul,
	year = {2025},
	note = {Publisher: Nature Publishing Group},
	pages = {6419},
}

@article{kudelin_photonic_2024,
	title = {Photonic chip-based low-noise microwave oscillator},
	volume = {627},
	copyright = {2024 The Author(s)},
	issn = {1476-4687},
	url = {https://www.nature.com/articles/s41586-024-07058-z},
	doi = {10.1038/s41586-024-07058-z},
	language = {en},
	number = {8004},
	urldate = {2025-07-23},
	journal = {Nature},
	author = {Kudelin, Igor and Groman, William and Ji, Qing-Xin and Guo, Joel and Kelleher, Megan L. and Lee, Dahyeon and Nakamura, Takuma and McLemore, Charles A. and Shirmohammadi, Pedram and Hanifi, Samin and Cheng, Haotian and Jin, Naijun and Wu, Lue and Halladay, Samuel and Luo, Yizhi and Dai, Zhaowei and Jin, Warren and Bai, Junwu and Liu, Yifan and Zhang, Wei and Xiang, Chao and Chang, Lin and Iltchenko, Vladimir and Miller, Owen and Matsko, Andrey and Bowers, Steven M. and Rakich, Peter T. and Campbell, Joe C. and Bowers, John E. and Vahala, Kerry J. and Quinlan, Franklyn and Diddams, Scott A.},
	month = mar,
	year = {2024},
	note = {Publisher: Nature Publishing Group},
	pages = {534--539},
}

@article{sun_integrated_2024,
	title = {Integrated optical frequency division for microwave and {mmWave} generation},
	volume = {627},
	copyright = {2024 The Author(s)},
	issn = {1476-4687},
	url = {https://www.nature.com/articles/s41586-024-07057-0},
	doi = {10.1038/s41586-024-07057-0},
	language = {en},
	number = {8004},
	urldate = {2025-07-23},
	journal = {Nature},
	author = {Sun, Shuman and Wang, Beichen and Liu, Kaikai and Harrington, Mark W. and Tabatabaei, Fatemehsadat and Liu, Ruxuan and Wang, Jiawei and Hanifi, Samin and Morgan, Jesse S. and Jahanbozorgi, Mandana and Yang, Zijiao and Bowers, Steven M. and Morton, Paul A. and Nelson, Karl D. and Beling, Andreas and Blumenthal, Daniel J. and Yi, Xu},
	month = mar,
	year = {2024},
	note = {Publisher: Nature Publishing Group},
	pages = {540--545},
}

@article{ludlow_optical_2015,
	title = {Optical atomic clocks},
	volume = {87},
	url = {https://link.aps.org/doi/10.1103/RevModPhys.87.637},
	doi = {10.1103/RevModPhys.87.637},
	number = {2},
	urldate = {2025-07-23},
	journal = {Reviews of Modern Physics},
	author = {Ludlow, Andrew D. and Boyd, Martin M. and Ye, Jun and Peik, E. and Schmidt, P. O.},
	month = jun,
	year = {2015},
	note = {Publisher: American Physical Society},
	pages = {637--701},
}

@article{xu_sensing_2019,
	title = {Sensing and tracking enhanced by quantum squeezing},
	volume = {7},
	copyright = {© 2019 Chinese Laser Press},
	issn = {2327-9125},
	url = {https://opg.optica.org/prj/abstract.cfm?uri=prj-7-6-A14},
	doi = {10.1364/PRJ.7.000A14},
	language = {EN},
	number = {6},
	urldate = {2025-07-23},
	journal = {Photonics Research},
	author = {Xu, Chuan and Zhang, Lidan and Huang, Songtao and Ma, Taxue and Liu, Fang and Yonezawa, Hidehiro and Zhang, Yong and Xiao, Min},
	month = jun,
	year = {2019},
	note = {Publisher: Optica Publishing Group},
	pages = {A14--A26},
}

@article{lihachev_low-noise_2022,
	title = {Low-noise frequency-agile photonic integrated lasers for coherent ranging},
	volume = {13},
	copyright = {2022 The Author(s)},
	issn = {2041-1723},
	url = {https://www.nature.com/articles/s41467-022-30911-6},
	doi = {10.1038/s41467-022-30911-6},
	language = {en},
	number = {1},
	urldate = {2025-07-23},
	journal = {Nature Communications},
	author = {Lihachev, Grigory and Riemensberger, Johann and Weng, Wenle and Liu, Junqiu and Tian, Hao and Siddharth, Anat and Snigirev, Viacheslav and Shadymov, Vladimir and Voloshin, Andrey and Wang, Rui Ning and He, Jijun and Bhave, Sunil A. and Kippenberg, Tobias J.},
	month = jun,
	year = {2022},
	note = {Publisher: Nature Publishing Group},
	pages = {3522},
}

@article{mecozzi_use_2023,
	title = {Use of {Optical} {Coherent} {Detection} for {Environmental} {Sensing}},
	volume = {41},
	issn = {1558-2213},
	url = {https://ieeexplore.ieee.org/document/10064176/},
	doi = {10.1109/JLT.2023.3252444},
	number = {11},
	urldate = {2025-07-23},
	journal = {Journal of Lightwave Technology},
	author = {Mecozzi, Antonio and Antonelli, Cristian and Mazur, Mikael and Fontaine, Nicolas and Chen, Haoshuo and Dallachiesa, Lauren and Ryf, Roland},
	month = jun,
	year = {2023},
	pages = {3350--3357},
}

@article{kikuchi_fundamentals_2016,
	title = {Fundamentals of {Coherent} {Optical} {Fiber} {Communications}},
	volume = {34},
	issn = {1558-2213},
	url = {https://ieeexplore.ieee.org/document/7174950},
	doi = {10.1109/JLT.2015.2463719},
	number = {1},
	urldate = {2025-07-23},
	journal = {Journal of Lightwave Technology},
	author = {Kikuchi, Kazuro},
	month = jan,
	year = {2016},
	pages = {157--179},
}

@article{liu_integrated_2024,
	title = {Integrated photonic molecule {Brillouin} laser with a high-power sub-100-{mHz} fundamental linewidth},
	volume = {49},
	copyright = {© 2023 Optica Publishing Group},
	issn = {1539-4794},
	url = {https://opg.optica.org/ol/abstract.cfm?uri=ol-49-1-45},
	doi = {10.1364/OL.503126},
	language = {EN},
	number = {1},
	urldate = {2024-09-19},
	journal = {Optics Letters},
	author = {Liu, Kaikai and Wang, Jiawei and Chauhan, Nitesh and Harrington, Mark W. and Nelson, Karl D. and Blumenthal, Daniel J.},
	month = jan,
	year = {2024},
	note = {Publisher: Optica Publishing Group},
	pages = {45--48},
}

@inproceedings{puckett_higher_2019,
	title = {Higher {Order} {Cascaded} {SBS} {Suppression} {Using} {Gratings} in a {Photonic} {Integrated} {Ring} {Resonator} {Laser}},
	url = {https://ieeexplore.ieee.org/document/8750460},
	doi = {10.1364/CLEO_SI.2019.SM4O.1},
	urldate = {2024-07-22},
	booktitle = {2019 {Conference} on {Lasers} and {Electro}-{Optics} ({CLEO})},
	author = {Puckett, Matthew and Bose, Debapam and Nelson, Karl and Blumenthal, Daniel J.},
	month = may,
	year = {2019},
	note = {ISSN: 2160-8989},
	pages = {1--2},
}

@article{kharel_noise_2016,
	title = {Noise and dynamics in forward {Brillouin} interactions},
	volume = {93},
	url = {https://link.aps.org/doi/10.1103/PhysRevA.93.063806},
	doi = {10.1103/PhysRevA.93.063806},
	number = {6},
	urldate = {2024-06-14},
	journal = {Physical Review A},
	author = {Kharel, P. and Behunin, R. O. and Renninger, W. H. and Rakich, P. T.},
	month = jun,
	year = {2016},
	note = {Publisher: American Physical Society},
	pages = {063806},
}

@article{behunin_fundamental_2018,
	title = {Fundamental noise dynamics in cascaded-order {Brillouin} lasers},
	volume = {98},
	doi = {10.1103/PhysRevA.98.023832},
	journal = {Physical Review A},
	author = {Behunin, Ryan and Otterstrom, Nils and Rakich, Peter and Gundavarapu, Sarat and Blumenthal, Daniel},
	month = feb,
	year = {2018},
}

\section{Appendix}
\appendix
\section{Correlation functions}
\begin{widetext}
With the solution for $\hat{b}_{jk}$ in \eqref{eq:bhatcorrelation} and the properties of the phonon Langevin force $\xi_{jk}$, we find the two time correlation functions
\begin{align}
    \langle\hat{b}_{jk}^\dag(t)\hat{b}_{j'k'}(t')\rangle=&\int_{-\infty}^td\tau\int_{-\infty}^{t'}d\tau'e^{(i\Delta\Omega_{jk}-\Gamma_{jk}/2)(t-\tau)}e^{-(i\Delta\Omega_{jk}+\Gamma_{jk}/2)(t'-\tau')}\delta_{jj'}\delta_{kk'}\Gamma_{jk}n_{th}\delta(\tau-\tau')e^{-i\Omega_{jk}'(\tau-\tau')}\nonumber\\&=n_{jk}\delta_{jj'}\delta_{kk'}e^{-\frac{\Gamma_{jk}}{2}|t-t'|}e^{i\Delta\Omega_{jk}(t-t')}
\end{align}
and 
\begin{align}
    \langle\hat{b}_{jk}(t)\hat{b}^\dag_{j'k'}(t')\rangle=(n_{jk}+1)\delta_{jj'}\delta_{kk'}e^{-\frac{\Gamma_{jk}}{2}|t-t'|}e^{-i\Delta\Omega_{jk}(t-t')}.
\end{align}
Using this to calculate the correlation functions for the optical fluctuating forces, $\tilde{h}_p$ and $\tilde{h}_l$ (above threshold), we find
\begin{align}
   \langle \tilde{h}_p^\dag(t)\tilde{h}_p(t')\rangle= &\langle\bigl[\eta^\dag_p(t)+i\sum_kg^*_{lk}\alpha_le^{-i\varphi_l(t)}\hat{b}_{lk}^\dag(t) \bigr]e^{i\varphi_p(t)}\times\bigl[\eta_p(t')-t\sum_{k'}g_{lk'}\alpha_le^{i\varphi_l(t')}\hat{b}_{lk'}(t')\bigr]e^{-i\varphi_p(t')}\rangle
   \\&=\langle\eta_p^\dag(t)\eta_p(t')\rangle\langle e^{-i\varphi_p(t)}e^{i\varphi_p(t')}\rangle+\sum_{k,k'}g_{lk}^*g_{lk'}\alpha_l^2\langle\hat{b}^\dag_{lk}(t)\hat{b}_{lk'}(t')\rangle\langle e^{-i[\varphi_l(t)-\varphi_p(t)]}e^{i[\varphi_l(t')-\varphi_p(t')]}\rangle\nonumber
\end{align}
Due to the assumption that fluctuation in phase is a white noise gaussian process, $\langle e^{-i\varphi_p(t)}e^{i\varphi_p(t')}\rangle=1$ and
\begin{align}
    \langle e^{-i[\varphi_l(t)-\varphi_p(t)]}e^{i[\varphi_l(t)-\varphi_p(t')]}\rangle\sim e^{-\frac{1}{2}\gamma_\phi|t-t'|},
    \label{eq:phase_correlation}
\end{align}
where $\gamma_\phi$ characterizes the correlation time of the phases. This will relate to the laser linewidths, which is much less than the phonon damping rate, i.e., $\gamma_\phi\ll\Gamma_{ph}$. In this limit, Eq. \eqref{eq:phase_correlation} $\sim1$. This gives
\begin{align}
    \langle\tilde{h}_p^\dag(t)\tilde{h}_p(t')\rangle\approx\gamma N_p\delta(t-t')+\sum_k|g_{lk}|^2\alpha_l^2\delta_{kk'}n_{lk}e^{-\frac{\Gamma_{lk}}{2}|t-t'|}e^{i\Delta\Omega_{lk}(t-t)}.
\end{align}
Following the same logic, we arrive at the two-time correlation function for $\tilde{h}_l$,
\begin{align}
    \langle\tilde{h}_l^\dag(t)\tilde{h}_l(t')\rangle\approx\gamma N_l\delta(t-t')+\sum_k|g_{lk}|^2\alpha_p^2\delta_{kk'}(n_{lk}+1)e^{-\frac{\Gamma_{lk}}{2}|t-t'|}e^{-i\Delta\Omega_{lk}(t-t)}.
\end{align}
For $\langle\tilde{h}_p(t)\tilde{h}_p^\dag(t')\rangle$, change $N_p\rightarrow N_p+1$, $n_{lk}\rightarrow n_{lk}+1$, and flip the phase of $e^{i\Delta\Omega_{lk}(t-t')}$. For $\langle\tilde{h}_l(t)\tilde{h}_l^\dag(t')\rangle$, change $N_l\rightarrow N_l+1$, $(n_{lk}+1)\rightarrow n_{lk}$, and flip the phase. Additionally, we find $\langle \tilde{h}_{p} (t) \tilde{h}_{p}(t') \rangle=\langle \tilde{h}_{l} (t) \tilde{h}_{l}(t') \rangle=0$ and
\begin{align}
    \langle \tilde{h}_p(t)\tilde{h}_l(t')\rangle=-\sum_k|g_{lk}|^2\alpha_p\alpha_l(n_{lk}+1)e^{-\frac{\Gamma_{lk}}{2}|t-t'|}e^{-i\Delta\Omega_{lk}(t-t)}=\langle \tilde{h}^\dag_l(t)\tilde{h}^\dag_p(t')\rangle.
\end{align}
where $p$ and $l$ can be exchanged by changing $n_{lk}+1\rightarrow n_{lk}$ and flipping the phase. Leveraging the definitions ${\rm Re}[h]=1/2(h+h^\dag)$ and ${\rm Im}[\tilde{h}]=-i/2(h-h^\dag)$, we compute additional useful correlation functions here,
\begin{align}
    \langle {\rm Re}&[\tilde{h}_{p/l}(t)]{\rm Re}[\tilde{h}_{p/l}(t')]\rangle=\langle {\rm Im}[\tilde{h}_{p/l}(t)]{\rm Im}[\tilde{h}_{p/l}(t')]\rangle\\&=\frac{1}{2}\gamma\delta(t-t')(N_{p/l}+1/2)+\frac{1}{4}\sum_k |g_{lk}|^2\alpha_{l/p}^2e^{-\frac{\Gamma_{lk}}{2}|t-t'|}\bigl[ n_{lk}e^{i\Delta\Omega_{lk}(t-t')}+(n_{lk}+1)e^{-i\Delta\Omega_{lk}(t-t')}\bigr]\nonumber
\end{align}
\begin{align}
    \langle {\rm Re}[\tilde{h}_{p}(t)] &{\rm Re}[\tilde{h}_{l}(t')]\rangle=\langle {\rm Re}[\tilde{h}_{l}(t)] {\rm Re}[\tilde{h}_{p}(t')]\rangle\\&=-\frac{1}{4}\sum_k|g_{lk}|^2\alpha_l\alpha_p\delta_{kk'}e^{-\frac{\Gamma_{lk}}{2}|t-t'|}\bigl[ n_{lk}e^{i\Delta\Omega_{lk}(t-t')}+(n_{lk}+1)e^{-i\Delta\Omega_{lk}(t-t')}\bigr]\nonumber
\end{align}
\begin{align}
    \langle {\rm Re}[\tilde{h}_{p}(t)] &{\rm Im}[\tilde{h}_{l}(t')]\rangle=-\langle {\rm Im}[\tilde{h}_{l}(t)] {\rm Re}[\tilde{h}_{p}(t')]\rangle\\&=\frac{1}{4i}\sum_k|g_{lk}|^2\alpha_l\alpha_pe^{-\frac{\Gamma_{lk}}{2}|t-t'|}\bigl[ n_{lk}e^{i\Delta\Omega_{lk}(t-t')}-(n_{lk}+1)e^{-i\Delta\Omega_{lk}(t-t')}\bigr]\nonumber.
\end{align}

\end{widetext}
\end{document}